\pdfoutput=1  % for arXiv pdflatex processing

\documentclass[10pt]{article}
\usepackage[margin=3.0cm]{geometry}
\usepackage{amsfonts,amsmath,amsthm,amssymb}
\usepackage{mathpazo}           % math & rm
\usepackage{bbold} 				% for identity operator/symbol
\usepackage[nottoc,numbib]{tocbibind} % ensure biblio appears in TOC
\usepackage{fancyhdr}
\usepackage[affil-sl]{authblk}  % authors and affiliations w/ maketitle
\usepackage{framed}             % for mybox environment
\usepackage{enumitem}           % Roman items and also referencing same
\usepackage{appendix}           % better section titles
\usepackage{graphicx}
\usepackage{caption}
\usepackage{subcaption}         % subfigure pkg replacement
\usepackage{verbatim}
\usepackage{colortbl}           % color row in Table 5
\usepackage[dvipsnames]{xcolor} % for CiteColor in hyperref
\usepackage[bookmarks=false,colorlinks=true,citecolor=JungleGreen,urlcolor=blue]{hyperref}
\usepackage[normalem]{ulem}	    % better underlining
\newlist{inlinelist}{enumerate*}{1}
\setlist*[inlinelist,1]{%
  label=(\roman*),
}

\newenvironment{mybox} % put box around important text
{ \vspace{10pt}\noindent\begin{minipage}[!hp]{\textwidth}\begin{framed}}
{ \end{framed}\end{minipage}\vspace{10pt}\noindent }

\newcommand{\gov}{{WEB.gov}}
\newcommand{\get}{{\tt HTTP GET}}
\newcommand{\post}{{\tt HTTP POST}}

\def\figs{Figures} 		% path to figure files

\newtheorem{exam}{Example}

\title{\sf How to Emulate Web Traffic Using Standard Load Testing Tools}

\author{James F. Brady}
\affil{\small State of Nevada, Carson City, NV 89701 \authorcr 
\href{jfbrady@admin.nv.gov}{\tt jfbrady@admin.nv.gov}} 

\author{Neil J. Gunther}
\affil{\small Performance Dynamics, Castro Valley, CA 94552 \authorcr 
\href{mailto:njgunther@perfdynamics.com}{\tt njgunther@perfdynamics.com}} 

\date{}

\begin{document}
\sffamily
\maketitle

\thispagestyle{empty} % no p.1 on 1st page
%% Special running head for preprints
%\thispagestyle{fancy} % Must go after maketitle to get header on p. 1
\lhead{\em CMG imPACt 2016}
\chead{}
%\rhead{\em \copyright~2016 Performance Dynamics}
\lfoot{} 
\cfoot{\thepage}
\rfoot{}
\pagestyle{fancy} % put page # in footer

\begin{abstract} \normalsize
Conventional load-testing tools are based on a fifty-year old time-share
computer paradigm where a finite number of users submit requests and respond in
a synchronized fashion. Conversely, modern web traffic is essentially
asynchronous and driven by an unknown number of users. This difference presents
a conundrum for testing the performance of modern web applications. Even when
the difference is recognized, performance engineers often introduce
modifications to their test scripts based on folklore or hearsay published in
various Internet fora,  much of which can lead to wrong results. We present a
coherent methodology,  based on two fundamental principles,  for emulating web
traffic using a standard load-test environment.
\end{abstract}

\newpage
\tableofcontents

% Methodology
% A system of broad principles or rules from which specific methods or procedures may be derived to interpret or solve different problems within the scope of a particular discipline. Unlike an algorithm, a methodology is not a formula but a set of practices.
% From http://www.businessdictionary.com/definition/methodology.html

% Principle of Operation
% Engineering explanation of how something works
% What we have is a Testing Principle

\section{Introduction}
Conventional load testing tools, e.g., \href{http://jmeter.apache.org}{Apache JMeter} and \href{http://www8.hp.com/us/en/software-solutions/loadrunner-load-testing/}{HP LoadRunner}, are simulators. More particularly, they are workload simulators that generate real-life application loads on a representative computing platform. These simulations are based on the paradigm that a fixed, finite, set of load-generator threads mimic the actions of a finite population of human users who submit requests to the test platform: the System Under Test (SUT). 
With the advent of web applications, however, this paradigm has shifted because the number of 
active users on the Internet is no longer fixed. At any point in time that number is dynamic and the workload mix is changing. This stands in stark contrast to the fifty-year old monolithic, time-share, computer paradigm inherent in standard load-testing tools.

Many performance engineers are polarized over how best to represent web-users with existing load-test tools. Some believe there is no significant issue, as long the number of load-generators is relatively large with the 
\hyperlink{gloss:ztime}{think time} often set to zero (see Sect.~\ref{sec:princopsA}), while others who recognize the problem are confused about how to correctly represent web requests~\cite{GOOG14}. 
%\marginpar{\footnotesize What about SPEC and TPC benchmark generators?}
It should be emphasized that this subject is fraught with confusion due to a lack of clear terminology and rigorous methods. See, e.g.,~\cite{GOOG14,HASH10,NING09,SCHR06}. 
%See also the confusion over the term {\em concurrency} in Section~\ref{sec:closed}.
We attempt to correct that in this paper by developing two new web testing principles. As far as we are aware, this is the first time a consistent framework for simulating web-users has been presented in the context of performance engineering. 

Our starting point is to recognize that typical test workloads and web workloads are very different
and therefore need to be formalized very differently.  
Next, we establish the formal criteria needed to generate web requests. 
We then show how to verify the load generators actually produce web-like requests during a test.  
Finally, we present a detailed case study based on our methodology. The approach is similar to that implemented by AT\&T Bell Laboratories to size telephone network components. There, fundamental queue-theoretic principles~\cite{GUNT11} formed the foundation and measurements were used to confirm it~\cite{HAYW82}. 
Our approach can be viewed as the historical descendent of the methods first devised  one hundred years ago by \href{https://en.wikipedia.org/wiki/Agner_Krarup_Erlang}{Agner Erlang} to do capacity planning 
for the telephone network belonging to the Copenhagen Telephone Company: the Internet of his day~\cite{ERLA17}.

This paper is organized as follows. 
Section~\ref{sec:workloads} establishes the difference between real web-based users and so-called {\em virtual users} in conventional load-test tools. 
Section~\ref{sec:methodology} presents the first of the principles that define our methodology, viz., web testing \hyperlink{box:princA}{Principle A}. 
Section~\ref{sec:application} presents the second of the principles that define our methodology, viz., web testing \hyperlink{box:princB}{Principle B}, and demonstrates the application of those two principles to load testing a large-scale website.
Section~\ref{sec:summary} summarizes our conclusions.  
The reader is encouraged to click on technical terms in the text that are linked to definitions in a Glossary (Appendix~\ref{sec:glossary}),
particularly because various authors employ many of the same terms differently from us---another potential source of confusion.

\section{Load Testing Simulations} \label{sec:workloads}
To commence from a very broad perspective, the key difference between the workload generated using conventional load-test tools and the workload generated by actual web-based users can be characterized as the difference in the pattern of requests arriving at the SUT. 
Virtual users in a conventional load-testing environments generate a {\em synchronous} arrival pattern, whereas  web-based users generate an {\em asynchronous} arrival pattern.  
Similarly, the queue-theoretic terms, \hyperlink{gloss:closed}{closed} system and \hyperlink{gloss:open}{open} system~\cite{GUNT11}, have  also been used  respectively in this connection~\cite{DIAO01,SCHR06}.
We show how all these terms are related in Section~\ref{sec:methodology}.

\subsection{Synchronous vs. asynchronous arrivals} \label{sec:async}
\begin{figure}[!ht]
    %\centering
    \begin{subfigure}[b]{0.4\textwidth}
        \includegraphics[scale = 0.6]{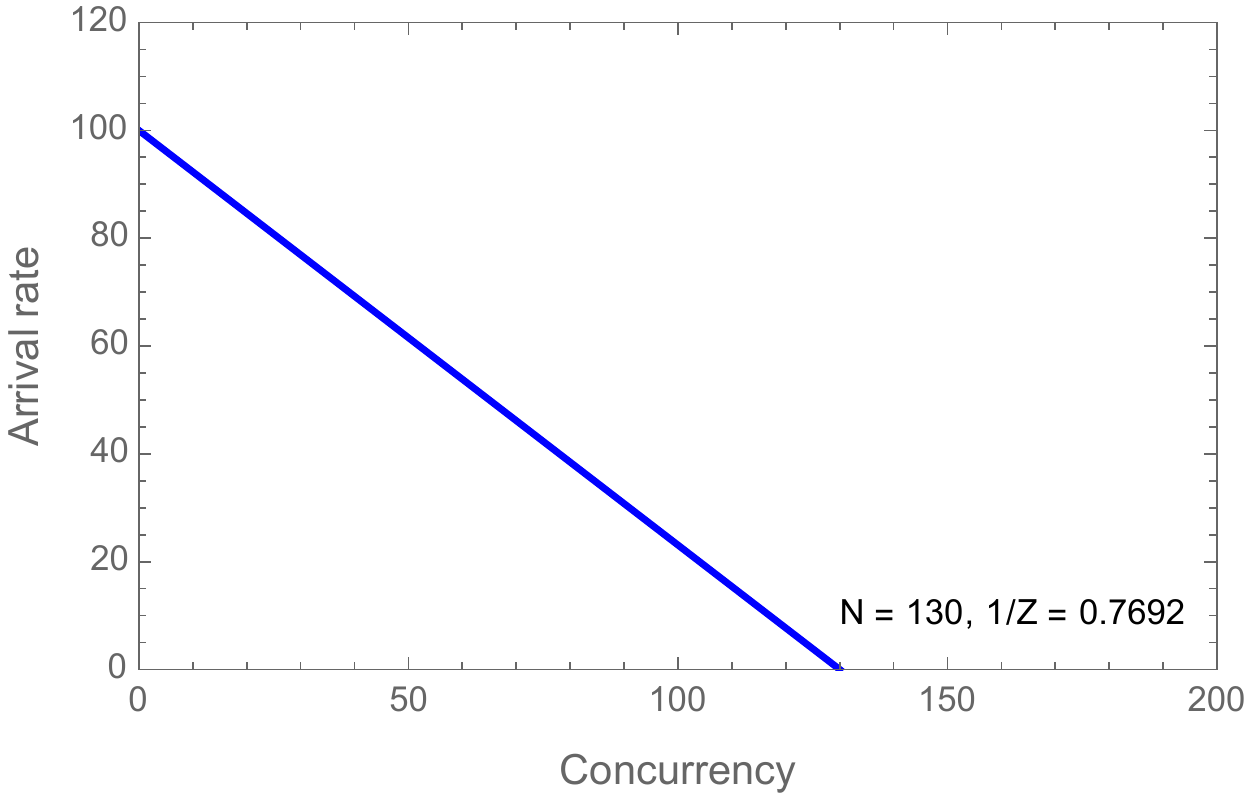}
        \caption{Synchronous (closed) central server}  \label{fig:lambda-closed}
    \end{subfigure}
    \hspace{1.5cm} 
    \begin{subfigure}[b]{0.4\textwidth}
        \includegraphics[scale = 0.6]{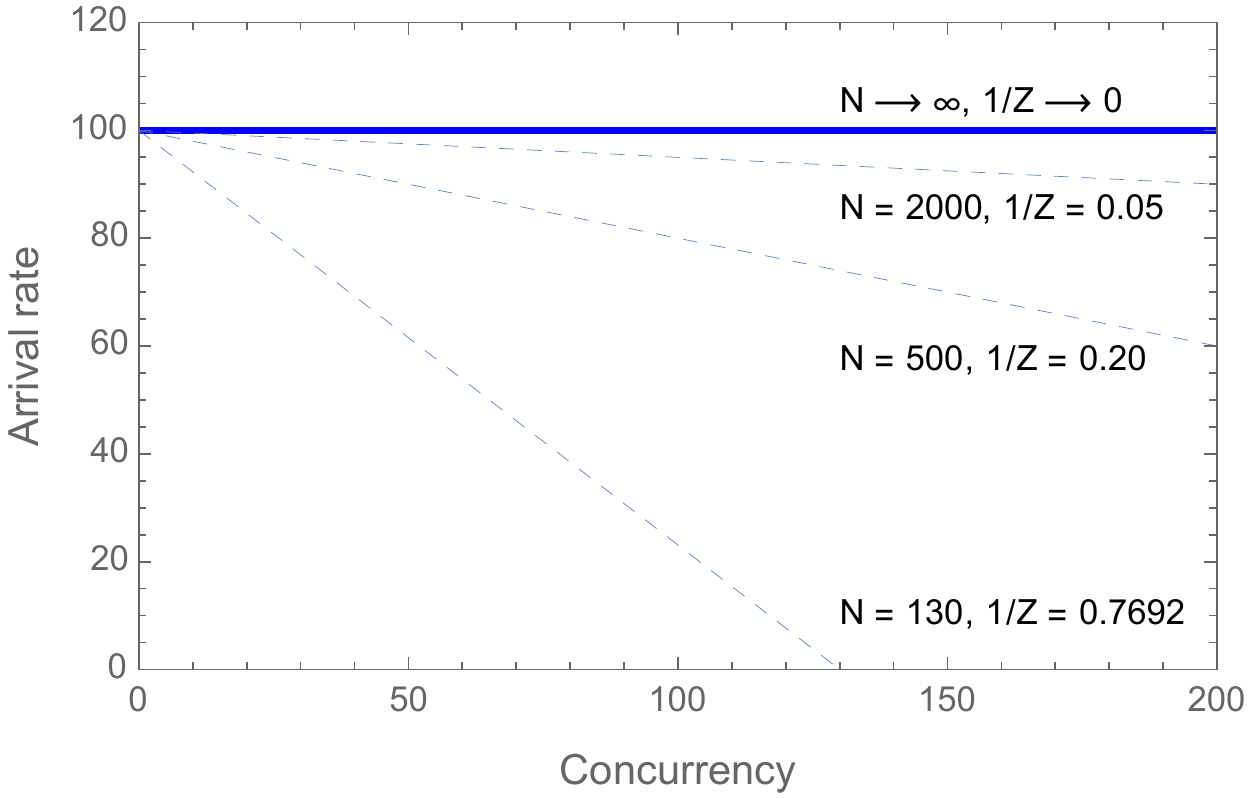}
        \caption{Asynchronous (open) web server}  \label{fig:lambda-open}
    \end{subfigure}
    \caption{Comparison of arrival rates as a function of concurrent requests}  \label{fig:lambdas}
\end{figure}

Figure~\ref{fig:lambdas} 
is a graphical representation of the difference between synchronous arrivals in a \hyperlink{gloss:closed}{closed} system and asynchronous arrivals in a \hyperlink{gloss:open}{open} system. 
Figure~\ref{fig:lambda-closed} shows how the average \hyperlink{gloss:arate}{arrival rate} 
into a typical (closed) test system {\em decreases} linearly with the level of \hyperlink{gloss:conc}{concurrency}  because the rate of requests arriving from the load generators depends {\em synchronously} on the queueing state in the SUT. 

Figure~\ref{fig:lambda-closed} represents the case of a load test system with up to $N = 130$ virtual users, each with a mean think time setting of $Z=1.3$ seconds. The maximum \hyperlink{gloss:arate}{arrival rate} of 100 \get s/second occurs at near-zero concurrency in the SUT, i.e., when there are few active virtual users. Conversely, at maximal load with all $N = 130$ virtual users active, the \hyperlink{gloss:arate}{arrival rate} at the SUT falls to zero. The level of \hyperlink{gloss:conc}{concurrency} in the SUT cannot be greater than the number of active load generators. The slope of the load line corresponds to $1/Z$: the mean request rate of each virtual user.

This retrograde effect---higher \hyperlink{gloss:conc}{concurrency} causes a lower request rate---seems counterintuitive but is easily explained. 
The logic in a typical load generation script  only allows one outstanding request at a time, i.e.,  each virtual user cannot issue its next request until the current request has completed and returned a result. In other words, 
the more outstanding requests there are in the SUT, the fewer new requests can be initiated by the load generators. In this way, the driver-side of the test environment is synchronously coupled to the state of the SUT 
and produces the self-throttling characteristic seen in Fig.~\ref{fig:lambda-closed}.
We shall revisit this effect in more detail in Section~\ref{sec:methodology}.

No such self-throttling is possible in a purely asynchronous or \hyperlink{gloss:open}{open} system.
To understand this difference consider the standard load test system in Fig.~\ref{fig:lambda-closed} where the number of active virtual users becomes extremely large ($N \rightarrow \infty$). 
This situation is depicted in Fig.~\ref{fig:lambda-open} with the load progressively increased as  
$N = 130, 500, 2000, \ldots$ virtual users. Notice that the slope of the corresponding load line 
also decreases (due to the decreasing value of $1/Z$) 
until it would ideally become constant at infinite load: the horizontal dashed line with zero slope in Fig.~\ref{fig:lambda-open}. 
In the infinite load limit there is no dependency between the arrival rate and the level of concurrency in the SUT. The \hyperlink{gloss:arate}{arrival rate} becomes completely asynchronous or decoupled from the state of the SUT.
A more detailed explanation is provided in Section~\ref{sec:princopsA}.

This is most like the situation in real web sites. 
Requests, such as an \get, arrive into the web server from what appears to be an infinite or indeterminate number of users on the Internet. The only quantities that can be measured directly are the time at which requests arrive into the TCP listen-queue and the number of the requests in each  measurement interval.
The arithmetic mean of those samples can then be used to calculate summary statistics, such as the 
average \hyperlink{gloss:arate}{arrival rate} of requests. 
The number of requests residing in the web server is a measure of the level of 
\hyperlink{gloss:conc}{concurrency} in the system and is largely independent of the \hyperlink{gloss:arate}{arrival rate}. 
Sect.~\ref{sec:closed} contains a more detailed discussion about \hyperlink{gloss:conc}{concurrency}.

The goal of our methodology is to find a way to reduce synchronization effects in the SUT so that requests appear to be generated asynchronously, or nearly so. 
Put more visually, we want to ``rotate'' the load line in Fig.~\ref{fig:lambda-closed} counterclockwise until its slope approaches the horizontal line in Fig.~\ref{fig:lambda-open}. Moreover, we want to achieve that result without necessitating any exotic modifications to the conventional load-test environment.

\subsection{Web simulation tools} \label{sec:tools}
To give some perspective for our later discussion, we provide a brief overview 
of other tools that claim to simulate web traffic. 
Commercial load testing products come with their own inherent constraints for web-scale workloads due to 
licensing fees being scaled to the number of cloned load generators. For example, 
a relatively modest 500 client LoadRunner license can cost around \$100,000~\cite{QUOR10}. 
Open source JMeter, however, can be extended to more than $1000$ users by resorting to a JMeter cluster to provide enough driver capacity. But that approach also comes with its own limitations~\cite{FOAL15}.
JMeter can also be configured to do \href{http://jmeter.apache.org/usermanual/jmeter_distributed_testing_step_by_step.pdf}{distributed testing}, including  on 
\href{http://stackoverflow.com/questions/16618915/setting-up-jmeter-for-distributed-testing-in-aws-with-connectivity-issues}{\em Amazon Web Services}.
Commercial web-based simulation tools are provided by 
\href{https://www.soasta.com/solutions/performance-testing/}{SOASTA}, \href{https://bmp.lightbody.net}{BrowserMob}, \href{https://www.neustar.biz/services/web-performance/load-testing}{Neustar}, and \href{https://loadimpact.com}{Load Impact}.
% wikipedia list https://en.wikipedia.org/wiki/Category:Load_testing_tools

Some cautionary remarks about these tools are in order. 
It can be confusing to understand whether tests are being run in their entirety on a cloud-based platform, like  AWS EC2 instances or, whether the EC2 instances are being used merely as load generators to drive the web site SUT across the Internet. In other words, the performance engineer needs to understand whether the load generators are {\em local} or {\em remote}, relative to the SUT.
Those questions notwithstanding, none to these approaches would make any difference if the scripts still produce synchronous arrivals as defined in Sect.~\ref{sec:async}.

\subsection{Virtual users and web users} \label{sec:users}
In the conventional language of load testing, the load generators represent virtualized human users or {\em virtual users}, for short. It is these virtual users that provide the test workload. 
In fact, virtual users are just a fixed set of process threads that run on separate computational resources that are distinct from the SUT.
As well as emitting \get\ and \post\ requests into the SUT, virtual-user threads typically incur a 
parameterized delay between receiving the response to the previous request and issuing the next request. 
Since this delay is intended to represent the time a human user spends ``thinking'' about what to do next, and therefore what kind of request is issued next, it is commonly known as the \hyperlink{gloss:ztime}{think time} and is often denoted by $Z$ in queue-theoretic parlance~\cite{GUNT11}.

The conventional load testing methodology has all the virtual-user threads that belong to a  
given group perform the same sequence of requests repeatedly for the entire period of each load test. 

\begin{figure}[!h]
    \centering
    \begin{subfigure}[b]{0.4\textwidth}
        \includegraphics[scale = 0.175]{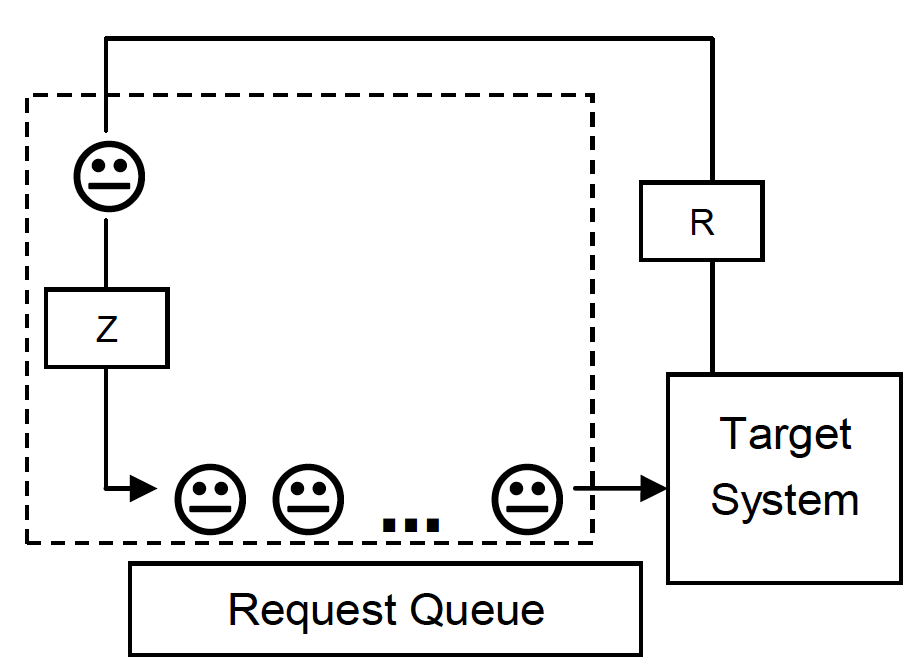} % for arXiv acceptance
        \caption{Virtual users}  \label{fig:frowns}
    \end{subfigure}
    \quad
    \begin{subfigure}[b]{0.4\textwidth}
        \includegraphics[scale = 0.25]{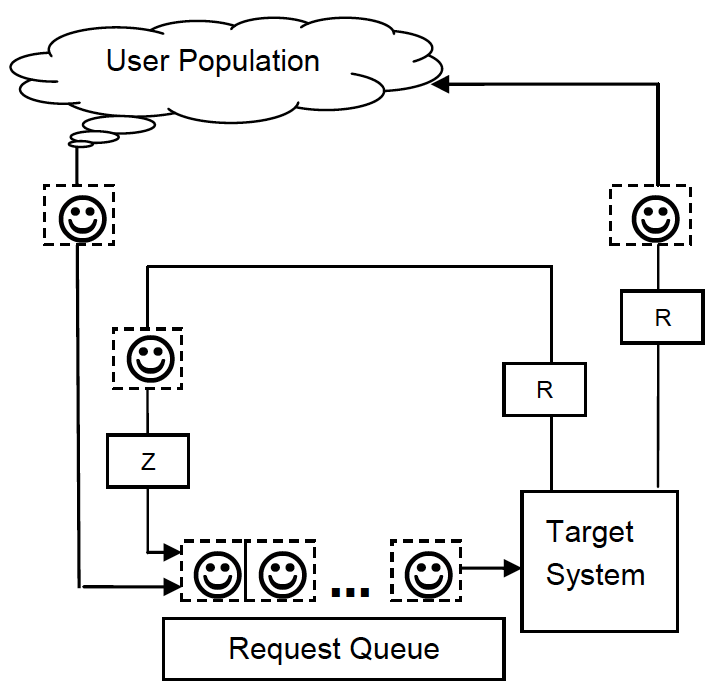} % for arXiv acceptance
        \caption{Web users}  \label{fig:smiles}
    \end{subfigure}
    \caption{Comparison of virtual users in a test rig and web users on the Internet}  \label{fig:smileys}
\end{figure}

Figure~\ref{fig:frowns} depicts this fixed virtual-user simulation schematically 
with each virtual-user thread, represented by a straight face 
\includegraphics[scale=0.025]{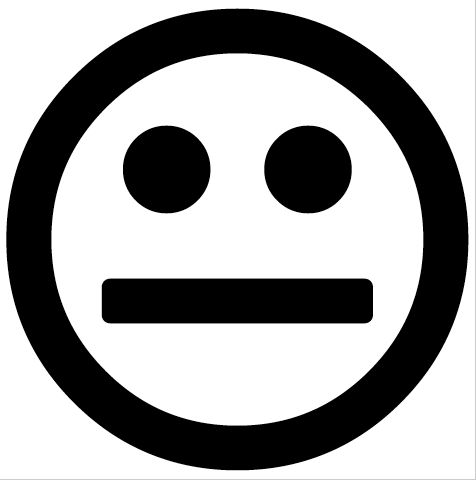},  
%\Neutrey[1.5][white],  
repeatedly cycling through five phases:
\begin{itemize} 
\item Make a web request into the target SUT
\item Wait for and measure the time of the \hyperlink{gloss:rtime}{response} (with mean time $R$)
\item Sleep for a \hyperlink{gloss:ztime}{think time} (with mean time $Z$)
\item Wake up and \hyperlink{gloss:wtime}{wait} in the OS run-queue (with mean time $W$)
\item Compute the requested results and return them (with mean \hyperlink{gloss:stime}{service time} $S$)
\end{itemize}

The activity of real web users, however, differ significantly from this scenario. 
Figure~\ref{fig:smiles} depicts these web users as smiley faces 
\includegraphics[scale=0.025]{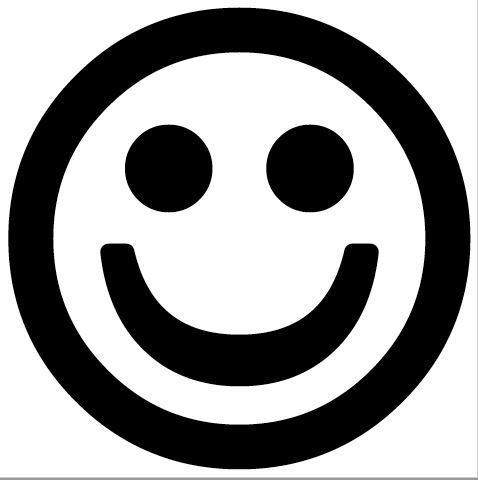}
%\Smiley[1.5][white].
Although Figs.~\ref{fig:frowns} and~\ref{fig:smiles} look similar, 
the following differences are important to understand.

\begin{description}
\item[\hypertarget{list:driveros}{Driver operating system:}]  
The dotted box in Fig.~\ref{fig:frowns} represents a single operating system instance. 
In a standard load-test environment, all the virtual users share that same operating system. 
Conversely, the real web-users in Fig.~\ref{fig:smiles} have a 
\includegraphics[scale=0.025]{\figs/smiley.png}
%\Smiley[1.5][white] 
face because each of them runs in their own private operating system.
In other words, the activity of a real user cannot be limited by an exhausted thread pool. 
Failure to recognize this potential driver-side constraint is a classic gottcha in load testing scenarios.
See, e.g.,~\cite{BUCH01} and~\cite[Chap. 12]{GUNT11}.

\item[\hypertarget{list:hetero}{Heterogeneous requests:}] 
Each 
\includegraphics[scale=0.025]{\figs/neutrey.png}
%\Neutrey[1.5][white] 
virtual user generally issues only one \get\ or \post\ request at a time.  
Each 
\includegraphics[scale=0.025]{\figs/smiley.png} 
%\Smiley[1.5][white] 
web user, on the other hand, can issue more than one type of HTTP request to   
retrieve multiple web objects that may comprise a complete web page.

\item[\hypertarget{list:cycling}{Cycling population:}] 
Real web users are not restricted by the closed-loop structure of the standard test environment in Fig.~\ref{fig:frowns}.  Rather, 
\includegraphics[scale=0.025]{\figs/smiley.png}
%\Smiley[1.5][white] 
web-users can come and go from a much larger but inactive population of users. This difference provides a pivotal insight for our approach. 
\end{description}

Overall, these distinctions tell us that real web users act like a variable sized set of threads that move in and out of the active thread pool to initiate HTTP requests. 
Accordingly, we want to find a way to have the fixed virtual users in Fig.~\ref{fig:frowns} 
emulate the real web users in Fig.~\ref{fig:smiles}.

\section{Web Testing Methodology} \label{sec:methodology} 
The key insight into producing an asynchronous request pattern with otherwise synchronous load-test tools 
comes from a comparison of the load-test simulation models with the appropriate queue-theoretic 
models~\cite{GUNT10a,GUNT10b,GUNT11,GUNT15,BRAD14,SCHR06}. 
Queueing metrics best capture the nonlinear relationship between well-known performance metrics. 
For our purposes, we only need to undertand two basic types of queues:
\begin{description}
\item [Open queue:] 
An \hyperlink{gloss:open}{open} queueing system involves asynchronous requests where the number of users generating those requests is unknown.
\item [Closed queue:] 
A \hyperlink{gloss:closed}{closed} queueing system involves synchronized requests where the number of users generating requests is finite and fixed.
\end{description}
The dynamics of these queues differ dramatically and therefore they need to be characterized in completely different ways.
We consider asynchronous request patterns first because they are associated with a simpler queueing structure than synchronous request patterns, which we deal with in Sect.~\ref{sec:closed}.

\subsection{Asynchronous open users} \label{sec:open}
\begin{figure}[!ht]
\centering
\includegraphics[scale = 0.65]{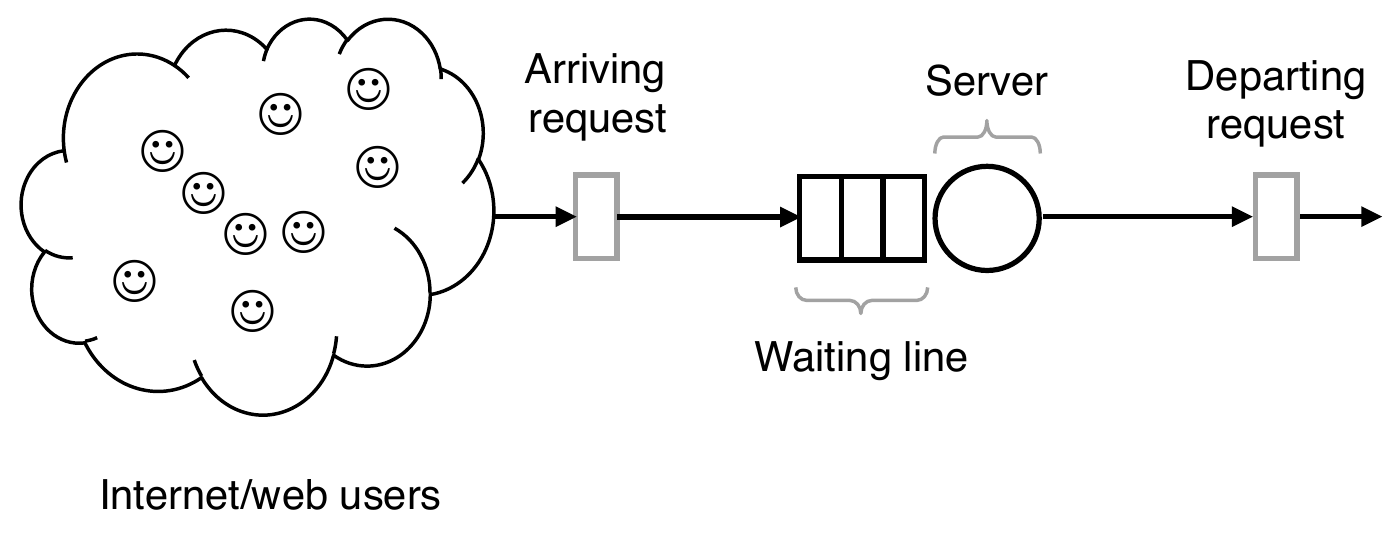} 
\caption{Open queue schematic representing asynchronous web requests} \label{fig:mm1} 
\end{figure}

The asynchronous workload referred to in Section~\ref{sec:workloads} is called an 
\hyperlink{gloss:open}{open} system in queueing theory parlance~\cite{GUNT11,KLEI75}. The term ``open'' is meant to indicate that requests arrive from an unknown number of users existing {\em outside} the queueing facility---indicated by the cloud in Fig.~\ref{fig:mm1}, which should be compared with Fig.~\ref{fig:smiles}.

An \hyperlink{gloss:open}{open} queueing center, shown schematically in  Fig.~\ref{fig:mm1}, is the simplest to characterize with   queueing metrics that represent averages or statistical means. Requests arrive from the left at a measurable rate $\lambda$, wait in line for some time, $W$, incur a time, $S$, to get serviced and finally depart the facility. 
The total time spent in the queueing facility is given by
\begin{equation}
R = W + S    \label{eqn:reztime}
\end{equation}
and is called the \hyperlink{gloss:reztime}{residence time}. It is the time spent in the queueing facility from arrival to departure.  $R$ can also be regarded as the \hyperlink{gloss:rtime}{response time} for a single queue.

This \hyperlink{gloss:open}{open} queue can be thought of as a very simple representation of a web site where the arrivals are 
\get\ requests generated by an unknown, and possibly large, number of Internet-based users. 
Some proportion of those users will be  actively sending requests to the web site while others are perusing the content resulting from previous web server responses. In queueing theory parlance, these latter users are said to be in a ``think state.''

Although the actual number of users submitting requests is indeterminate, 
the {\em number of requests resident in the web server}, $Q$, is simply the sum of those requests that are waiting for service ($\lambda W $) plus those that are already in service ($\lambda S$). From eqn.~\eqref{eqn:reztime}:
\begin{equation}
Q = \lambda W + \lambda S = \lambda R \label{eqn:qlength}
\end{equation}
This is one version of Little's law~\cite{GUNT14b}.
The steady-state \hyperlink{gloss:arate}{arrival rate} can be measured directly with a probe or calculated using the 
definition~\cite{GUNT11} 
\begin{equation}
\lambda=A/T .  \label{eqn:arate}
\end{equation}
where $A$ is the counted number of arrivals and $T$ is measurement period. 
The corresponding departure rate is $X=C/T$, where $C$ a count of the number of completed requests. 
During any suitably long measurement period, the \hyperlink{gloss:open}{open} queue will be in {\em steady state} such that arrivals count will closely match the number of completed requests, i.e., $A=C$.
It follows from this that $\lambda = X$ in steady state. The departure rate is more commonly known as the 
{\em throughput} and is a system metric that is measured by all load test tools. 

In many practical situations, the periods between arriving requests are found to be exponentially 
distributed~\cite{BRAD12}. See also, {\em Are Your Data Poissonian?} in~\cite[Chap. 3]{GUNT11}.  When the service periods are also exponentially distributed with mean time, $S$, the queue in Fig.~\ref{fig:mm1} is denoted M/M/1, where `M' denotes ``memoryless'' (or Markovian, more formally). 
This memoryless feature is a property of the \hyperlink{gloss:exp}{exponential distribution} 
%(see Glossary~\ref{sec:glossexp}) 
which, in turn, is associated with a \hyperlink{gloss:pproc}{Poisson process}. 
%(see Glossary~\ref{sec:glosspproc}).
Essentially, it means that history is no guide to the future when the time periods are 
exponentially distributed.
We will make use of this feature in Sect.~\ref{sec:application}.

Our use of the term {\em asynchronous} also seems particularly appropriate in light of Erlang's original development of the M/M/m queue to predict the waiting times at telephone switches. The `A' in ATM stands for `asynchronous' and {\em Asynchronous Transfer Mode} switches form the backbone of the modern Internet.

The important point for our discussion is that the explicit relationship between the asynchronous \hyperlink{gloss:arate}{arrival rate}, $\lambda$, and the  number of active Internet users is unknown. Consequently, there can be an unbounded (but not infinite) number of requests in the system, independent of $\lambda$. (See Fig.~\ref{fig:lambda-open})

\subsection{Synchronous closed users} \label{sec:closed}
As noted earlier, virtual users in a standard load-test system \hyperlink{list:cycling}{cycle} through the SUT 
in the sense that no new request can be issued by a virtual user until the currently outstanding request in the SUT has been serviced.
%(See {\em Cycling population} in  Sect.~\ref{sec:users})
This loop structure depicted in Fig.~\ref{fig:frowns} will now be reflected in a corresponding 
\hyperlink{gloss:closed}{closed} queueing model.

\begin{figure}[!ht]
\centering
\includegraphics[scale = 0.65]{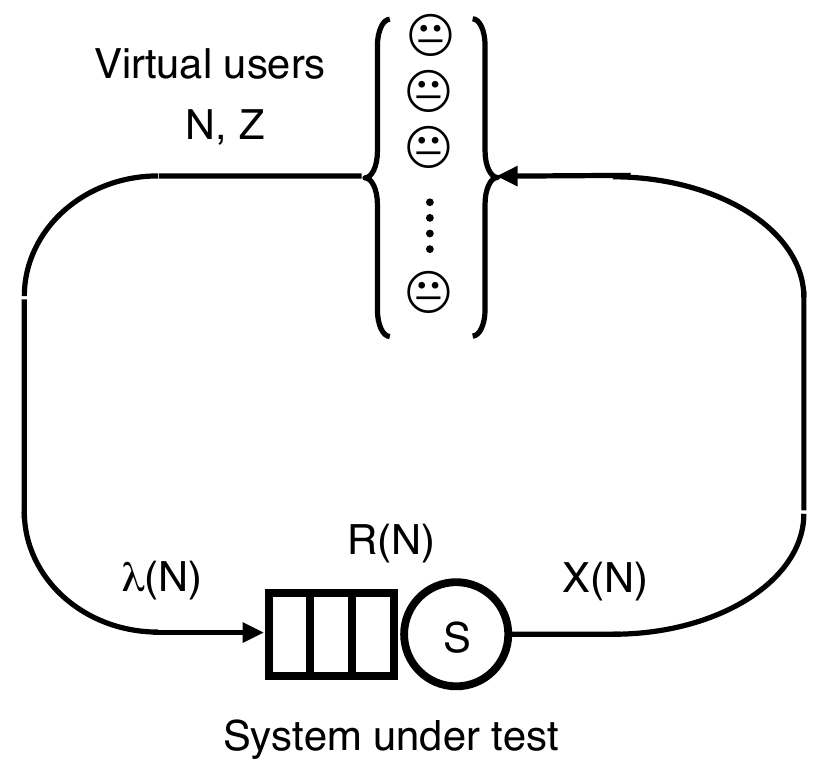} 
\caption{Closed queue schematic representing synchronous load-test requests} \label{fig:mm1n}
\end{figure}

In contrast to an \hyperlink{gloss:open}{open} queue, a \hyperlink{gloss:closed}{closed}  queue is actually comprised of two queues connected by the flow of requests and responses circulating between them. 
For historical reasons in the development of queueing theory, this system is sometimes referred to as the {\em central server} or the {\em repairman} model~\cite{GUNT11}. The two queues are:
\begin{enumerate}
\item 
$N$ sources of requests ({\em top queue} in Fig.~\ref{fig:mm1n}). These sources correspond to the 
load-test generators representing virtual users. Since each source effectively has its own service center (shown as 
\includegraphics[scale=0.025]{\figs/neutrey.png}
%\Neutrey[1.5][white] 
faces corresponding to those in Fig.~\ref{fig:frowns}), no waiting line forms at this queue.
\item 
The system under test ({\em bottom queue} in Fig.~\ref{fig:mm1n}). 
Although there are potentially many queueing facilities associated with actual computing resources in the SUT, for simplicity and without loss of generality, we show the SUT here as a single queue.
\end{enumerate}

\noindent
Since there are no requests arriving from outside this system (the meaning of ``closed''), the flow in Fig.~\ref{fig:mm1n} appears as a loop. 
 
For a given load test, each of $N$ generators executes a script that issues a request into the SUT. Unlike the \hyperlink{gloss:open}{open} queue of Sect.~\ref{sec:open}, with a potentially unbounded number of requests in the system, there can never be more than $N$ requests in the \hyperlink{gloss:closed}{closed} system.
The corresponding queue-theoretic notation is M/M/1/N/N~\cite{GUNT11}.
In actual test environments, the number of active generators may be restricted by 
the cost of licensing fees for commercial test tools, and 
the number of available threads. (See Sections~\ref{sec:gov} and~\ref{sec:pool})

The response of the SUT is measured in terms of the overall throughput, $X$ and 
\hyperlink{gloss:reztime}{residence time}, $R$. 
When the response is received by the associated generator, it will generally incur a predetermined delay, $Z$, before issuing the next request. This behavior is identical to the perusal time associated with the human users in the \hyperlink{gloss:open}{open} queue and is called the \hyperlink{gloss:ztime}{think time}. The average \hyperlink{gloss:ztime}{think time}, $Z$, is a programmable parameter in the generator script. During test execution, each script draws a random think-time variate from a known statistical distribution, e.g., a uniform or exponential distribution having a mean equal to $Z$.

A key distinction between an \hyperlink{gloss:open}{open} queue and a \hyperlink{gloss:closed}{closed} queue is that there cannot be more than one request outstanding from each generator. 
This constraint has the consequence that if all $N$ generators have issued requests into the SUT, there cannot be any  more arrivals until at least one of those requests has completed service, 
and also possibly incurred a think delay, $Z$. 
The effective \hyperlink{gloss:arate}{arrival rate} has fallen to zero!

In this sense, a \hyperlink{gloss:closed}{closed} queue is {\em self-throttling} due to the negative feedback loop in Fig.~\ref{fig:mm1n}. The more requests in the SUT, the lower the instantaneous \hyperlink{gloss:arate}{arrival rate} at the SUT, and vice versa. Moreover, the mean \hyperlink{gloss:arate}{arrival rate} is no longer constant but must therefore depend functionally on the $N$ users.  We present function, $\lambda(N)$ in 
eqn.~\eqref{eqn:crate} of Sect.~\ref{sec:princopsA}.

The total \hyperlink{gloss:rttime}{round-trip time}, $R_{TT}$, for a request in Fig.~\ref{fig:mm1n} is the sum of the time spent in the SUT and the time spent ``thinking'' in the generator script:
\begin{equation}
R_{TT} = R + Z \label{eqn:rtt}
\end{equation}
When a serviced request returns to its corresponding generator script, the \hyperlink{gloss:ztime}{think time}, $Z$, can be regarded as a special kind of 
\hyperlink{gloss:stime}{service time}---the service time on the driver side of the test platform (top queue in Fig.~\ref{fig:mm1n}). 
Then, $\lambda Z$ represents the total number of generators that are in a think state. 

Similarly, the total number of requests in the system is given by Little's law~\cite{GUNT11}
\begin{equation}
N = \lambda \, R_{TT}    \label{eqn:sysnum}
\end{equation}
which is the counterpart of eqn.~\eqref{eqn:qlength}. % was q
Expanding eqn.~\eqref{eqn:sysnum} produces
\begin{equation}
N = \lambda R + \lambda Z  \label{eqn:closedn} % was n
\end{equation}
In steady state, the total number of requests active in the \hyperlink{gloss:closed}{closed} system is the sum of those in the SUT, i.e., $\lambda R$, together with those in a think state, $\lambda Z$, on the generator side of the testing environment. We will make use of eqn.~\eqref{eqn:closedn} to determine the 
number of real web-users in Sect.~\ref{sec:ausers}. 
Substituting the expression for $Q$ from eqn.~\eqref{eqn:qlength} into eqn.~\eqref{eqn:closedn} shows how the total number of requests in the test system is related to the queue length 
(i.e., requests that are waiting together with requests that are already in service) on the SUT.
\begin{equation}
N = Q + \lambda Z \label{eqn:closednq} % was nq
\end{equation}

Equation~\eqref{eqn:closednq} provides an opportunity to disambiguate the term ``concurrency'' in the context of a load-test environment (see e.g.,~\cite{GOOG14,WARR13}). 
Broadly speaking, the level of \href{https://en.wikipedia.org/wiki/Concurrency_(computer_science)}{concurrency} refers to the number of processes or threads that are {\em acting} simultaneously. But, like the term {\em latency}, \hyperlink{gloss:conc}{concurrency} is often too generic to be really helpful for accurate performance analysis. For example, it could refer to:
\begin{enumerate}
\item the size of the generator thread-pool
\item the number of active threads
\item the number of virtual users
\item the number of active processes on the SUT
\end{enumerate}
This is where our notation is helpful.
We can immediately reduce the options to choosing between $N$ and $Q$. The difference between these quantities is the number of processes or threads that are in a think state, viz., $\lambda Z$. 
We know that $N$ represents the maximal number of users that can possibly generate requests into the SUT. 
In a \hyperlink{gloss:closed}{closed} queueing system, this number is split between 
the number of outstanding requests, $Q$, that are actively residing on the SUT and the number of 
 users that are thinking, $\lambda Z$. 
We note that \cite{COOP84} calls these thinking processes, ``idle'' with respect to the SUT in that they 
will eventually issue requests but have not done so, yet. 
Clearly, if all the generators were in such an idle state then, $Q=0$ and no load would be impressed on the SUT. 
If we take ``acting'' to mean executing on the computational resources of the SUT then, the level of \hyperlink{gloss:conc}{concurrency} is $Q$ in eqn.~\eqref{eqn:closednq}.
Indeed, $Q$ is the metric represented on the x-axis in Fig.~\ref{fig:lambdas}.

\subsection{Web testing principle A} \label{sec:princopsA}
Rearranging eqn.~\eqref{eqn:closednq} and dividing both sides by $Z$ yields
\begin{equation}
%\boxed{\lambda = \dfrac{N}{Z} - \dfrac{Q}{Z}}  \label{eqn:crate}
\lambda(N,Z) = \dfrac{N}{Z} - \dfrac{Q(N)}{Z} \label{eqn:crate}
\end{equation}
which tells us how the \hyperlink{gloss:arate}{arrival rate} varies as a function of both choice for the user load $N$ and the think time $Z$ setting in a \hyperlink{gloss:closed}{closed} system. 
For any chosen value of $Z$, eqn.~\eqref{eqn:crate} remains a nonlinear function of $N$
because the concurrency, $Q$, in the SUT is a nonlinear function of $N$. (See Fig.~\ref{fig:NZlimit})

The \hyperlink{gloss:arate}{arrival rate}, $\lambda(N,Z)$ in eqn.~\eqref{eqn:crate}, forms the basis of the methodology to be presented in the subsequent sections and 
to illustrate how it contains the performance characteristics of both virtual user and web user workloads, we developed a theoretical model in PDQ~\cite{GUNT15} the results of which are shown in Table~\ref{tab:znpdq}.

\begin{table}[!htb]
    \caption{Theoretical comparison of (a) constant Z and (b) scaled Z load models} \label{tab:znpdq}
    \small 
    \begin{subtable}{.5\linewidth} 
      \centering
        \caption{Constant $Z$ virtual users}  \label{tab:zconst}
        \begin{tabular}{rrrrrr}
		\bf N & \bf Z & \bf N/Z & \boldmath $\lambda$ & \bf  Q & \bf Q/Z\\
		\hline
          100 & 10 & 	10.00 & 	  9.80 & 	    2.01 & 	0.20\\
          200 & 10 & 	20.00 & 	 18.46 & 	   15.37 & 	1.54\\
          400 & 10 & 	40.00 & 	 19.95 & 	  200.52 & 	20.05\\
          600 & 10 & 	60.00 & 	 19.98 & 	  400.17 & 	40.02\\
          800 & 10 & 	80.00 & 	 19.99 & 	  600.09 & 	60.01\\
         1000 & 10 & 	100.00 & 	 19.99 & 	  800.05 & 	80.01\\
		\hline
	\end{tabular}
    \end{subtable}%
    \begin{subtable}{.5\linewidth}
      \centering
        \caption{Scaled $Z$ web users} \label{tab:zscaled}
		\begin{tabular}{rrrrrr}
		\bf N & \bf Z & \bf N/Z & \boldmath $\lambda$ & \bf  Q & \bf Q/Z\\
		\hline    
          100 &  5 &	20.00 &	 17.80 &	   11.02 &	2.20\\
          200 & 10 &	20.00 &	 18.46 &	   15.37 &	1.54\\
          400 & 20 &	20.00 &	 18.93 &	   21.38 &	1.07\\
          600 & 30 &	20.00 &	 19.14 &	   25.95 &	0.86\\
          800 & 40 &	20.00 &	 19.26 &	   29.78 &	0.74\\
         1000 &	50 &	20.00 &	 19.34 &	   33.15 &	0.66\\
		\hline
		\end{tabular}     
    \end{subtable}%
\end{table}
%%%
%%%\marginpar{\footnotesize cf. different $S_{max}$ in both cases}

Table~\ref{tab:zconst} corresponds to a typical load-test environment where the \hyperlink{gloss:ztime}{think time} is held constant at $Z=30$ seconds while the number of virtual users is increased by almost an order of magnitude in an attempt to emulate web users. Naturally, as $N$ increases, the ratio $N/Z$ increases. 
Similarly, the number of \hyperlink{gloss:conc}{concurrent} requests, $Q$, in the SUT also increases dramatically due to the  increasing number of load generators. 
Hence, the $Q/Z$ ratio increases. Recall that the synchronization of a \hyperlink{gloss:closed}{closed} system comes from the requirement that each virtual user can have no more than one outstanding request, and $Q$ is a measure of the total number of outstanding requests.
The difference between these two ratios in eqn.~\eqref{eqn:crate} is the average \hyperlink{gloss:arate}{arrival rate} into the SUT, $\lambda$, which clearly is not constant.
This already suggests that increasing $N$ alone cannot accurately mimic web traffic.

Table~\ref{tab:zscaled} shows what happens when the $N/Z$ ratio is held constant at 20 \get s/second. This is achieved by scaling the \hyperlink{gloss:ztime}{think time}, $Z$ by the same proportion as the increase in the number of generators, $N$. Notice the dramatic difference in the SUT \hyperlink{gloss:conc}{concurrency}. Compared with $Q$ in Table~\ref{tab:zconst}, it grows relatively slowly. This effect is explained as follows. 
Although the number of load generators is increased, as in the constant-$Z$ case, the delay between the completion of one request and the start of the next is also increased in the same proportion. Consequently, the number of \hyperlink{gloss:conc}{concurrent} requests in the SUT increases only very slowly. 
In this scaled-$Z$ case, the contribution from the $Q/Z$ ratio diminishes rapidly such that the effective \hyperlink{gloss:arate}{arrival rate} into the SUT quickly approaches the constant ratio $N/Z$. 
(See Fig.~\ref{fig:NZlimit})

%%% soft bottleneck and lambda_{infty} to go here
That both scenarios in Table~\ref{tab:znpdq} exhibit saturation effects is noteworthy because each is due to a very different cause.
The plateauing in $\lambda$ of Table~\ref{tab:zconst}, starting around $N=800$ virtual users, 
is the result of physical resource saturation in the SUT. 
It is a physical  bottleneck determined by the longest of the service demand, $S_{max}$, such that~\cite[Chap. 7]{GUNT11}
\begin{equation}
\lambda_{sat} = \dfrac{1}{S_{max}}  \label{eqn:xsat}
\end{equation}
The bound in eqn.~\eqref{eqn:xsat} represents the maximum possible throughput that the SUT can achieve.
In this case, $S_{max} = 50$ milliseconds so, $\lambda_{sat} = 20$  \get s/second.
In contrast, the plateauing in Table~\ref{tab:zscaled} is a result of constraining $N$ and $Z$ to scale as a fixed ratio
\begin{equation}
\lambda_{rat} = \dfrac{N}{Z}  \label{eqn:xrat}
\end{equation}
Since $N$ and $Z$ can be chosen arbitrarily, $\lambda_{rat}$ can take on any value as long as it satisfies the condition
\begin{equation}
\lambda_{rat} \leq \lambda_{sat}
\end{equation}
as exhibited in column 4 of Table~\ref{tab:zscaled}. 
The choice of $\lambda_{rat} = \lambda_{sat}$ in Table~\ref{tab:znpdq}  was made for simplicity.
In general, however, since $\lambda_{rat}$ can be chosen by the performance engineer, we shall refer to it as a {\em soft bottleneck}.
This throughput bound is depicted as a ({\em blue}) horizontal line in Fig.~\ref{fig:NZlimit} 
of Sect.~\ref{sec:convergence}.

Asynchronous requests, as defined in Sect.~\ref{sec:open}, create a constant mean \hyperlink{gloss:arate}{arrival rate} that  corresponds to the horizontal line in Fig.~\ref{fig:lambda-open}.
In that case, the goal should be to have $\lambda = \lambda_{rat}$ to reflect the \hyperlink{gloss:indpt}{independence} of the request rate from $Q$ in eqn.~\eqref{eqn:crate}. 
%(See Glossary~\ref{sec:glossindpt}) 
However, the presence of the $Q/Z$ term means $\lambda$ is not independent of $Q$ and that is responsible for the slope in the load line seen in Fig.~\ref{fig:lambda-closed}. 
If the magnitude of $Q$ could be made relatively small, that slope would become shallower which      
 would indicate the synchronization of virtual-user requests had become weaker. 
In other words, the synchronous requests  would look more and more like the 
asynchronous requests of the web users we want to mimic: the load line with near-zero slope in Fig.~\ref{fig:lambda-open}. 
The most efficient way to achieve that result is shown numerically in Table~\ref{tab:zscaled}. 
There, the ratio $N/Z$ is held constant while increasing the load $N$.
It is the {\em royal road} to producing asynchronous requests with \hyperlink{gloss:arate}{arrival rate} $\lambda_{rat}$. 
But how is this trick actually achieved in a real test environment?

Equation~\eqref{eqn:crate} assumes that the level of \hyperlink{gloss:conc}{concurrency} is smaller than the number of generators, i.e., $Q < N$. If it were the case that $Q > N$, the \hyperlink{gloss:arate}{arrival rate} at the SUT would be negative, which is not physically possible. 
If both parameters were very close in value, i.e., $Q \simeq N$, the \hyperlink{gloss:arate}{arrival rate} would become impractically small for load testing purposes. In other words, it is the relative magnitudes of each of $N, Q$, and $Z$ that determine the value of $\lambda$.

To help ensure $Q < N$ (actually, much smaller, $Q \ll N$, as we shall see) in a real test environment, the value of $Z$ can be {\em increased} in the test script.
In Table~\ref{tab:zconst}, each virtual user thinks for a mean period of $Z = 30$ time units. Suppose, however, that instead of treating $Z$ as a \hyperlink{gloss:ztime}{think time}, we treat it as free parameter in the load-test simulator, i.e., use $Z$ as a control knob to tune the web-user approximation to whatever accuracy we desire. 
In that vein, it is more useful to consider the inverse quantity, $1/Z$: the mean emission rate of user requests. 
If we retard the virtual user emission-rate by increasing $Z$, the number of \hyperlink{gloss:conc}{concurrent} requests in the SUT will also be reduced, i.e., $Q$ will become small. 
This effect can be seen clearly in Table~\ref{tab:zscaled} and a real example is presented in 
Sect~\ref{sec:gov}.

Now, this is where the magic happens. 
Little's law in eqn.~\eqref{eqn:qlength} tells us that the \hyperlink{gloss:rtime}{response time}, $R$, is directly proportional to the number of requests, $Q$, residing in the SUT. 
If $Q$ is long, then an arriving request will have to reside in the SUT for a longer time than if $Q$ were short. 
The size of $Q$ is determined by how many other requests are still outstanding in the SUT. 
Therefore, the expected $R$ of an arriving request from the $n$th virtual user 
depends on the state of the other $n-1$ virtual users. This dependency or interaction between users occurs in the queues of the SUT. In this way, virtual users are correlated in time and therefore lack 
statistical \hyperlink{gloss:indpt}{independence}. 
%(See Glossary~\ref{sec:glossindpt}) 
As $Q$ becomes small,  so does the time spent in the SUT. If $R$ is small (e.g., $R < Z$) then, the time for which each virtual user request remains outstanding also becomes small.
In the limit, if $Q=0$ or, equivalently $R=0$, there would be no outstanding requests at all (there would also be no work for the SUT to do) and each virtual user would no longer be constrained by the synchronization of the \hyperlink{gloss:closed}{closed} feedback loop in Fig.~\ref{fig:mm1n}.  Moreover, there would be no interaction in the SUT, so virtual users would also become statistically independent.
In other words, in that limit, requests would become asynchronous and virtual users would be transformed into statistically independent web users, each having a very low request rate, $1/Z$, because $Z$ is now relatively large. 
Since there are $N$ of these web users making requests, 
\begin{equation}
\lambda = \underbrace{ \dfrac{1}{Z} + \dfrac{1}{Z} + \ldots + \dfrac{1}{Z} }_{N~\text{web users}} = \dfrac{N}{Z},  \label{eqn:zsum}
\end{equation}
which is their aggregate request rate into the SUT. And since $\lambda$ would be independent of $Q$, it would also be consistent with Fig.~\ref{fig:lambda-open}.

Real web users, however, do induce work in the SUT, so the idealized requirement $Q=0$ is too severe.  
In fact, eqn.~\eqref{eqn:crate} only requires that the ratio $Q/Z$ be small, i.e., $Q$ can be non-zero as long as $Z$ is relatively large. This more relaxed condition means $N$ can also be relatively large without causing request synchronization to re-emerge, and that is what we see in  Table~\ref{tab:zscaled}.

\noindent
With this intuitive explanation of how virtual users are transformed into web users by eqn.~\eqref{eqn:crate}, we can state the first guiding principle of our test methodology. 

%%%%%%% Principle A %%%%%%%%%%
\begin{mybox} 
\begin{center} \hypertarget{box:princA}{\bf Principle A}\end{center}
To efficiently emulate web traffic the mean think-time, $Z$, in each of the $N$ load generators should 
be scaled with $N$ such that the ratio $N/Z$ remains constant 
thereby ensuring the request rate $\lambda$ approaches the soft bottleneck rate $\lambda_{rat}$.  
\end{mybox}
%%%%%%%%%%%%%%%%%

\hyperlink{box:princA}{Principle A} is the royal road for attaining \hyperlink{gloss:indpt}{statistical independence} by reducing the per-user request rate into the SUT. It also makes clear why the common practice among performance engineers of setting $Z=0$ in the generator scripts is actually fallacious when it comes to 
emulating web traffic. 
The intuition behind setting zero think time is that it 
increases the mean request rate into the SUT and that corresponds to the much larger number of real web users.
In fact, it is counterproductive for emulating web users, 
where the goal is to {\em reduce} the per-user request rate so as to keep $R$ small in the SUT relative to  $Z$. 
That is how the requirement of \hyperlink{gloss:indpt}{statistically independent} web requests is achieved.

We can also use \hyperlink{box:princA}{Principle A} to explain Fig.~\ref{fig:lambdas}.
In the synchronous queueing system of Fig.~\ref{fig:lambda-closed}, for a given load value $N$, 
the level of \hyperlink{gloss:conc}{concurrency}, $Q$, controls the \hyperlink{gloss:arate}{arrival rate} via the second term in 
eqn.~\eqref{eqn:crate}. 
Larger $Q$ value means longer queues so, the time spent in the SUT will also be longer. 
That also means requests remain outstanding for longer so, fewer new requests can be generated. 
That is the basis for the synchronous feedback loop in Figs.~\ref{fig:mm1n} and~\ref{fig:frowns}.
That the synchronous \hyperlink{gloss:arate}{arrival rate} decreases in direct proportion to $Q$ is reflected by the minus sign in eqn.~\eqref{eqn:crate}.
The negative sign is responsible for the negative slope in Fig.~\ref{fig:lambda-closed}. 

Conversely, in the asynchronous queueing system of Fig.~\ref{fig:lambda-open}, the \hyperlink{gloss:arate}{arrival rate} is independent of $Q$ and therefore appears as a horizontal line at height $N/Z$ on the y-axis.
This situation is tantamount to setting $Q=0$ in the second term of eqn.~\eqref{eqn:crate}, as already explained above.
The interested reader will find a different, but related, argument based on taking the continuum limit of a \hyperlink{gloss:bin}{Binomial distribution} in~\cite{COOP84}. 

Although we have used very simple queueing models in this section to establish 
\hyperlink{box:princA}{Principle A} of our methodology, the same results hold for more elaborate queueing models that reflect realistic systems.
For example, a multi-tier web site can be modeled by replacing the single queue representing the SUT in 
Fig.~\ref{fig:mm1n} by a set of single queues arranged in series and parallel with respect to their flows~\cite{GUNT11,GUNT15}.

\subsection{Visualizing principle A} \label{sec:convergence}
\begin{figure}[!h]
\centering
\includegraphics[scale = 0.75]{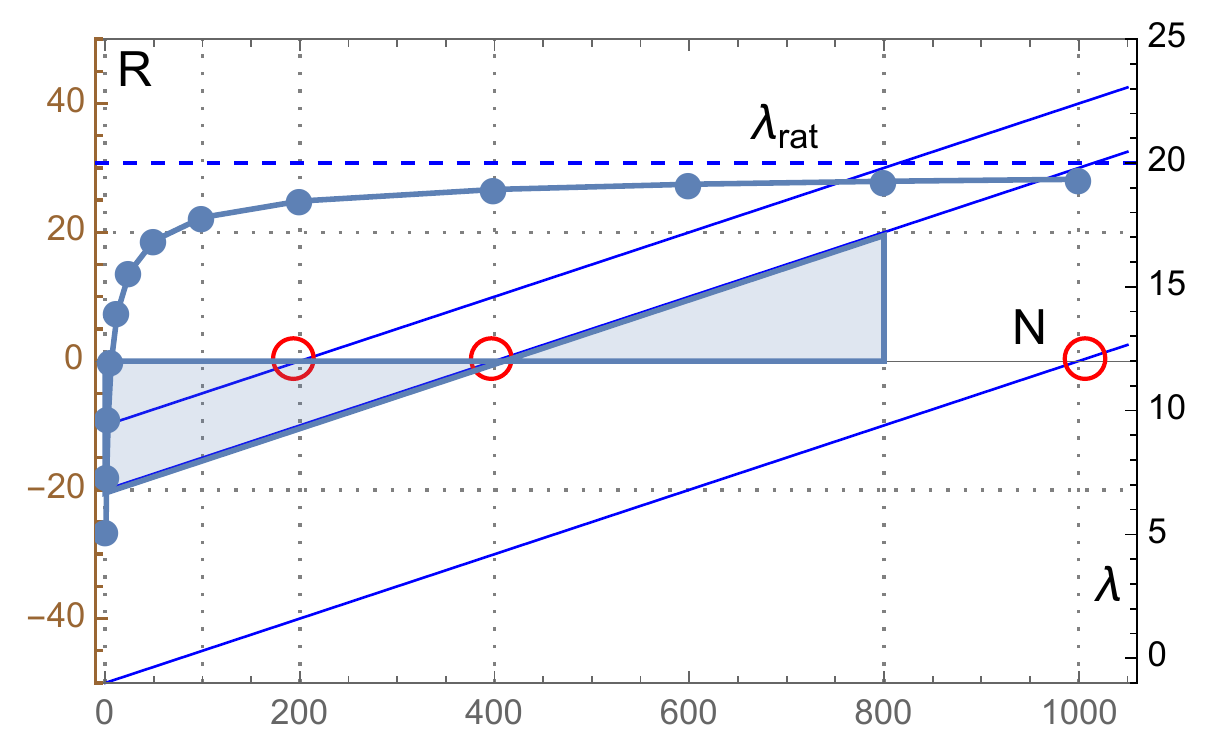} 
\caption{Convergence of $\lambda(N)$ toward $\lambda_{rat} = 20$ (triangle slope) according to Principle A} \label{fig:NZlimit}
\end{figure}

Figure~\ref{fig:NZlimit} presents a visual interpretation of eqn.~\eqref{eqn:crate} that underlies 
\hyperlink{box:princA}{Principle A}. 
The theoretical curve is the \hyperlink{gloss:arate}{arrival rate}, $\lambda(N)$ for any continuous value of $N$,
corresponding to the discrete load points in Table~\ref{tab:zscaled}, which are indicated by the blue dots 
sitting on that curve.

$\lambda(N)$ increases monotonically with $N$ and approaches the bound $\lambda_{rat}$ in eqn.~\eqref{eqn:xrat}, depicted by the {\em dashed}  horizontal line.
In this case, the soft-bottleneck is located at $\lambda_{rat} = 20$ \get s per second, as can be read off from the right-hand vertical axis. Notice that the plot actually has four axes:
\begin{enumerate}
\item 
A conventional $x$-axis (positioned as a {\em midline}) showing generator load in the range
$N \in [0,1000]$ 
\item 
A left-side $y$-axis  showing both positive \hyperlink{gloss:rtime}{response time} values  
in the range $R \in [0, +50]$ above the $x$-axis and negative values $R \in [0, -50]$ below the $x$-axis
\item 
A right-side $y$-axis showing positive \hyperlink{gloss:arate}{arrival rate} values $\lambda \in [0,25]$ 
\item 
Three diagonal lines passing through red circles at $N=200, 400, 1000$ that also intersect the 
left-side $y$-axis at $R=-10,-20,-50$, respectively
\end{enumerate}
Therefore, to fully appreciate the dense information contained in Fig.~\ref{fig:NZlimit} requires a little orientation. 

The continuous curve $\lambda(N)$ is computed for all possible theoretical load levels in the range $N \in (0, 1000]$ and therefore includes the specific data points of Table~\ref{tab:zscaled}. 
However, the curve is not determined uniquely by $N$. 
Equation~\eqref{eqn:crate} is really a function of both $N$ and $Z$, i.e., $\lambda = \lambda(N,Z)$, which  
simply  expresses the fact that the test workload intensity is controlled by the number of load generators ($N$) and the think delay ($Z$) set in the test script. Normally, $Z$ is a fixed constant so, there is no need to mention it explicitly. \hyperlink{box:princA}{Principle A}, however, requires that we do pay close attention to the $Z$ value because it is not fixed in eqn.~\eqref{eqn:crate}.

The $x$-position of each  data point ({\em solid dot}) corresponds to its load value $N$. 
The subset of data values in Table~\ref{tab:zscaled} corresponds to the six dots starting on the left side of the plot at $N = 100$ and ending on the right side at $N=1000$. 
To explain the procedure for finding the associated $Z$ values from the diagonals, we consider the  data points at $N = 200, 400$ and $1000$ virtual users as examples.

\begin{exam}[$N=200$ ratio] \label{exam:n200}
Consider the solid dot corresponding to $\lambda(200,Z)$ located at $N=200$. 
Identify the red circle below it and locate the diagonal that passes through that circle. 
Follow that diagonal downward to where it intersects the $y$-axis on the left side of Fig.~\ref{fig:NZlimit}.
The magnitude of that number corresponds to the associated $Z$ value. 
In this case, the diagonal intersects in the negative part of the $y$-axis (below the $x$-axis) at the value 
$-R = 10$. 
Since by definition the $y$-axis is the vertical line located at $x = 0$, the $R$-axis is located at $N = 0$ in  Fig.~\ref{fig:NZlimit}.
With zero active users, there are no requests in circulation and the 
\hyperlink{gloss:rttime}{round trip time} is $R_{TT} = 0$ at zero load.
By eqn.~\eqref{eqn:rtt} that is equivalent to $R+Z=0$.
Consequently, if $R = -10$ then $Z=10$, and the ratio $N/Z= 200/10 = 20$ \get s / second. 
\end{exam}

Note that the magnitude of the ratio $N/Z= 20$ is equivalent to the value of $\lambda_{rat}$ (the horizontal blue line) as indicated on the $\lambda$-axis of Fig.~\ref{fig:NZlimit}. 
It is the maximum possible arrival rate that can be achieved in this system that is consistent with the soft bottleneck defined by \hyperlink{box:princA}{Principle A}.

\begin{exam}[$N=400$ ratio] \label{exam:n400}
Repeating the same steps as Example~\ref{exam:n200} for the solid dot corresponding to $\lambda(400,Z)$ located at $N=400$, we see that the associated diagonal line intersects at $R = -20$ or $Z=20$. Again, 
\mbox{$N/Z= 400/20$} or equivalently, $\lambda_{rat} = 20$ \get s / second.
\end{exam}

\begin{exam}[$N=1000$ ratio] \label{exam:n1000}
Similarly, for the solid dot corresponding to $\lambda(1000,Z)$ located at $N=1000$ the associated diagonal intersects at $R = -50$ yielding  
\mbox{$N/Z= 1000/50$} or $\lambda_{rat} = 20$ \get s / second.
\end{exam}

As noted in Sect.~\ref{sec:princopsA}, 
the sharp nonlinearity in the curve arises from the rapid diminishment of the queueing term, $Q/Z$, in eqn.~\eqref{eqn:crate}. 
The difference between the two terms in eqn.~\eqref{eqn:crate} can also be visualized using  
Fig.~\ref{fig:NZlimit}. 
The difference is the vertical gap between the curve and the soft bottleneck bound at $\lambda_{rat}$ with  
the gap size equal to the magnitude of $Q/Z$. As $N$ is increased (to the right), the size of the gap gets smaller, which shows why $\lambda_{rat}$ is an asymptotic bound. 
The following examples provide numerical verification.

\begin{exam}[Vertical difference $\Delta\lambda(200)$]
From Example~\ref{exam:n200} we have $\lambda_{rat} = 20$ \get s/second and 
from Table~\ref{tab:zscaled} we note that \mbox{$Q/Z = 1.54$} at $N=200$ (row 2). 
The vertical difference between the horizontal line and the solid dot at $N=200$ is given by 
$\Delta \lambda = 20 - 1.54 = 18.46$.
This is precisely the height of the dot on the right-hand $\lambda$-axis, i.e., 
$\lambda(200,10) = 18.46$ \get s/second, in agreement with column 4 of Table~\ref{tab:zscaled}.
\end{exam}

\begin{exam}[Vertical difference $\Delta\lambda(400)$]
From Table~\ref{tab:zscaled} we have \mbox{$Q/Z = 1.07$} at $N=400$ (row 3). 
The vertical difference between the horizontal line and the solid dot at $N=400$ is given by 
$\Delta \lambda = 20 - 1.07 = 18.93$ 
or $\lambda(400,20) = 18.93$ \get s/second, in agreement with Table~\ref{tab:zscaled}.
\end{exam}

\begin{exam}[Vertical difference $\Delta\lambda(1000)$]
From Table~\ref{tab:zscaled} we have \mbox{$Q/Z = 0.66$} at $N=1000$ (row 6). 
The vertical difference is given by $\Delta \lambda = 20 - 0.66 = 19.34$ 
or $\lambda(1000,50) = 19.34$ \get s/second, as seen in Table~\ref{tab:zscaled}.
\end{exam}

Referring now to the shaded triangles in Fig.~\ref{fig:NZlimit}, the associated red circle at $N=400$ sits at the vertex of an upside-down right-angled triangle ({\em left shaded area}). 
The base of this triangle therefore has length of 400. 
 Its hypotenuse lies on the diagonal that intersects the $R$-axis at $R=-20$, as mentioned in Example~\ref{exam:n400}.
This just corresponds to an upside-down height $|R| = 20$. 
The conjugate right-angled triangle with positive height $R = +20$ is shown between the grid lines at 
$N=400~\text{and}~800$ ({\em right shaded area}). The  hypotenuse of both triangles lie on the common diagonal line. 
The slope of the diagonal line is given by the rise over the run of either triangle, viz., 
$Z/N = 20/400$, which is the inverse of $\lambda_{rat}$ analogous to $\lambda_{sat}$ in eqn.~\eqref{eqn:xsat}.
%to the slope $S_{max}$ of the hockey-stick handle in Fig.~\ref{fig:rt-asymptote}.

The same logic can be applied to the smaller  triangle formed by the first red circle at $N=200$, as well as the  much larger triangle formed by the rightmost red circle at $N=1000$.
Because all such triangles are \href{http://www.mathopenref.com/similartriangles.html}{similar triangles}, they 
explain why the ratio $N/Z=20$ is held constant in eqn.~\eqref{eqn:crate}.
Scaling the ratio $N/Z$ in \hyperlink{box:princA}{Principle A} is the same as maintaining the triangular  
proportions associated with each value of $N$ in Fig.~\ref{fig:NZlimit}. 

\begin{figure}[!h]
\centering
\includegraphics[scale = 0.45]{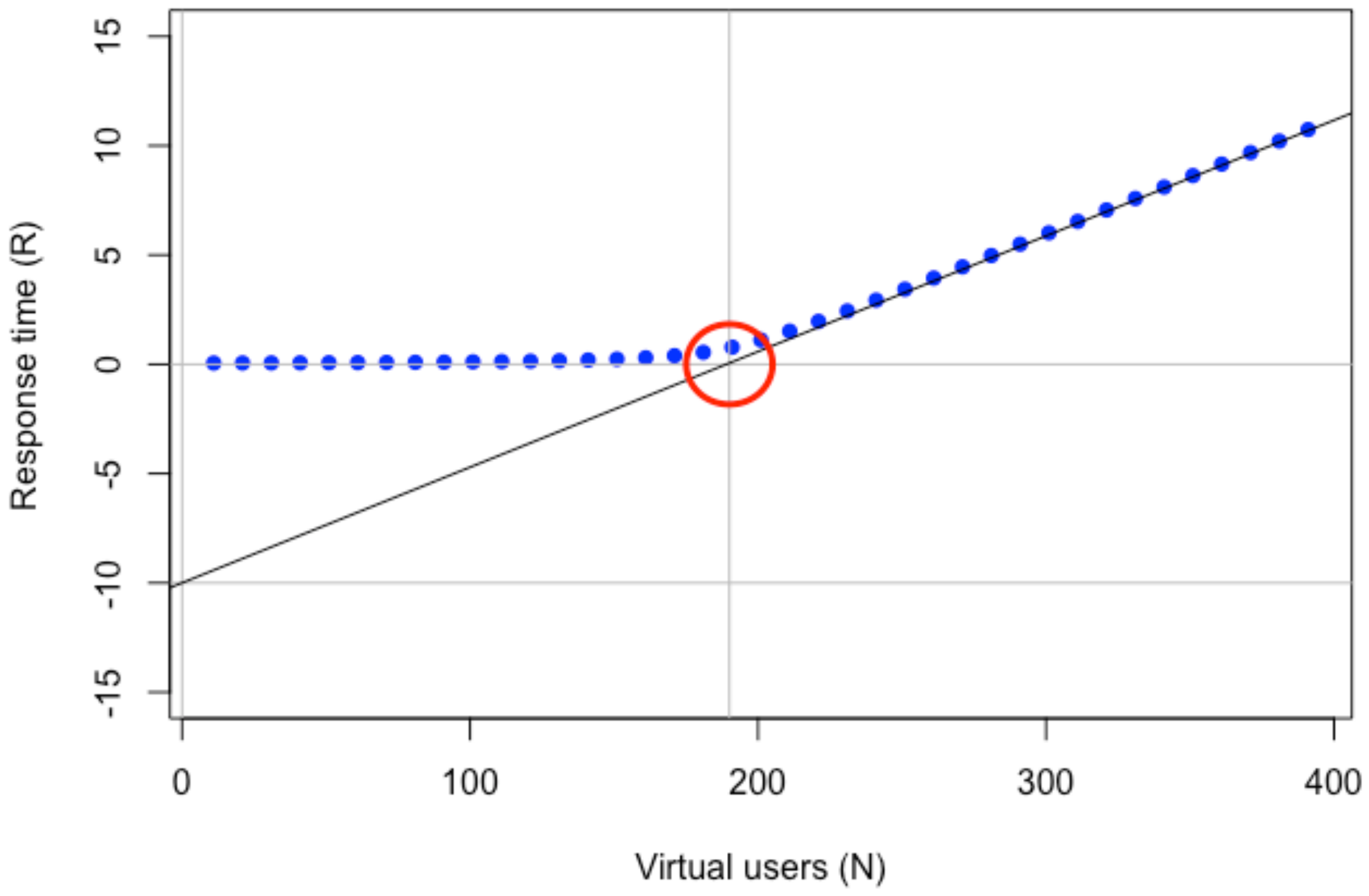} 
\caption{\small Typical response-time data ({\em solid dots}) for a closed queueing system with $Z=10$ showing the classic hockey-stick shape. The hockey-stick handle is always asymptotic to the diagonal line ($R_{\infty}$) given by eqn.~\eqref{eqn:rtasymp} with slope $S_{max}$. Extending the asymptote below the x-axis (horizontal line at $R=0$) always intersects the y-axis (vertical line at $N=0$) at $R=-Z$} \label{fig:rt-asymptote}
\end{figure}

Of course, the reader is likely wondering where the diagonal lines come from in the first place.
In \hyperlink{gloss:closed}{closed} queueing models, 
the  \hyperlink{gloss:ztime}{think time}, $Z$ is usually considered as part of the \hyperlink{gloss:rttime}{round-trip time}, $R_{TT}$ in eqn.~\eqref{eqn:rtt}, not the arrival rate, $\lambda$. However, 
in a load-test model  (Fig.~\ref{fig:mm1n}), $N$ and $Z$ are the two parameters on the driver side that control the rate of requests associated with the synchronous workload.
The diagonal lines in  Fig.~\ref{fig:NZlimit} belong to the response-time ($R$) 
for a \hyperlink{gloss:closed}{closed} queueing system: previously identified with the load test environment in 
Sect.~\ref{sec:closed}. 
Each diagonal line is the asymptote that controls the slope of the classic ``hockey stick'' shape of 
\hyperlink{gloss:rtime}{response time} measurements (cf. Fig.~\ref{fig:jimfig12} top-left) in a \hyperlink{gloss:closed}{closed} system~\cite[see Fig. 7.3 there]{GUNT11}. 
Figure~\ref{fig:rt-asymptote} is provided as a reminder of how this works.

The response-time data ({\em solid dots}) in Fig.~\ref{fig:rt-asymptote} start out approximately horizontally (i.e., roughly parallel to the x-axis) because the queue lengths are typically small at low loads, so the 
\hyperlink{gloss:rtime}{response times} are approximately constant or slowly increasing. 
The general expression for the \hyperlink{gloss:rtime}{response time} in a closed queueing system is given by 
\begin{equation}
R(N) = \dfrac{N}{X(N)} - Z   \label{eqn:rtime}
\end{equation}
The explicit loop structure in Fig.~\ref{fig:mm1n} is contained implicitly in eqn.~\eqref{eqn:rtime} since both the \hyperlink{gloss:rtime}{response time} $R$ and the throughput $X$ are dependent on $N$ so, in general, this equation cannot be solved analytically and one must resort to computational tools like PDQ~\cite{GUNT11}. 

Eventually, the bottleneck resource (i.e., the resource with the longest mean service period, $S_{max}$) saturates in accordance with eqn.~\eqref{eqn:xsat} and that causes queues to grow dramatically. 
This rapid increase in queue length (due to the number of \hyperlink{gloss:conc}{concurrent} requests in the SUT) causes the response-time data ({\em solid dots}) to increase along the diagonal ``handle'' of the hockey stick.
This follows from the fact that the asymptotic form of eqn.~\eqref{eqn:rtime}, i.e., for large $N$, is given 
by~\cite[Sect. 7.4]{GUNT11}
\begin{equation}
R_{\infty} = N \, S_{max} - Z  \label{eqn:rtasymp}
\end{equation}
This is simply the equation for a straight line with the general form $y=mx+c$ where $m$ is the gradient of the line and the constant $c$ is the point of intersection with the $y$-axis, i.e., the value of $y$ at $x=0$.
By analogy, the slope $m=S_{max}$ and 
the intersection with the $R$-axis  occurs at $N=0$ in eqn.~\eqref{eqn:rtasymp}, 
i.e., at $R_{\infty} = -Z$. In Fig.~\ref{fig:rt-asymptote} it corresponds to $R = -10$ or $Z=10$ at $N=0$.
This is the justification for introducing the diagonal axes or $Z$-lines in Fig.~\ref{fig:NZlimit} to indicate the associated value of $Z$ required by eqn.~\eqref{eqn:crate}.

Figure~\ref{fig:NZlimit} shows why $N$ or $Z$ must be varied as a {\em pair} of values $(N,Z)$ 
in order to mimic the asynchronous workload in Table~\ref{tab:zscaled}. 
The typical load-testing scenario in Table~\ref{tab:zconst}, with $N \rightarrow \infty$, is tantamount to 
letting the SUT choose $\lambda_{rat}$ for you, viz., $\lambda_{rat} = \lambda_{sat}$, which might not be 
appropriate.
Conversely, the scaled-$Z$ approach offers the performance engineer an infinite variety of choices for running 
load-test simulations with $\lambda_{rat}$ matched to measured values of real-world asynchronous web traffic.
We make use of similar opportunities in Sect.~\ref{sec:pool}.

\section{Applying the Methodology} \label{sec:application} 
Sect.~\ref{sec:princopsA} has provided the conceptual basis of the methodology 
embodied in \hyperlink{box:princA}{Principle A}. It tells us how to adjust the \hyperlink{gloss:ztime}{think time} setting in the load generators  
so that synchronous virtual users are transformed into asynchronous web users.
In this section, we want to verify that \hyperlink{box:princA}{Principle A} indeed produces the correct statistical traffic pattern associated with asynchronous web users. 
This is a necessary step because all the quantities in eqn.~\eqref{eqn:crate} are relative magnitudes, not absolutes, so we always need to check that we have not inadvertently chosen an incompatible set of magnitudes for $N,Q$, and $Z$.
This step will become \hyperlink{box:princB}{Principle B}, which we will then apply to testing a real website in Sect.~\ref{sec:princopsB}.  

What kind of statistical test should be contained in \hyperlink{box:princB}{Principle B}?
Here, we can look to A. K. Erlang as a role model. 
Erlang's first published work~\cite{ERLA09} was mostly about measuring telephone traffic and demonstrating that phone calls   arrived randomly into the switch (a female operator) according to a 
\hyperlink{gloss:pproc}{Poisson process}. (See Fig.~\ref{fig:stochprocs})
%(See Glossary~\ref{sec:glosspproc} 
Applying probability theory to an engineering problem was considered a novel idea a hundred years ago. 
%\marginpar{\footnotesize How did Erlang verify? $CoV$?} Not even close. Pure theory.

Following Erlang, we ask, do random web-user requests arriving into a web site also follow a \hyperlink{gloss:pproc}{Poisson process}? 
To some extent the answer to this question has already been pre-empted by the queue-theoretic analysis in 
Sect.~\ref{sec:open}. 
The memoryless property of Markovian queues is an attribute of a \hyperlink{gloss:pproc}{Poisson process}. 
%(See Glossary~\ref{sec:glosspproc})
Since we already know that a \hyperlink{gloss:pproc}{Poisson process} is the correct statistical model for web users, we just need a way to validate it in the context of load test measurements.

\subsection{Poisson properties}  \label{sec:poissonprops}
\begin{figure}[!h]
\centering
\includegraphics[scale = 0.20]{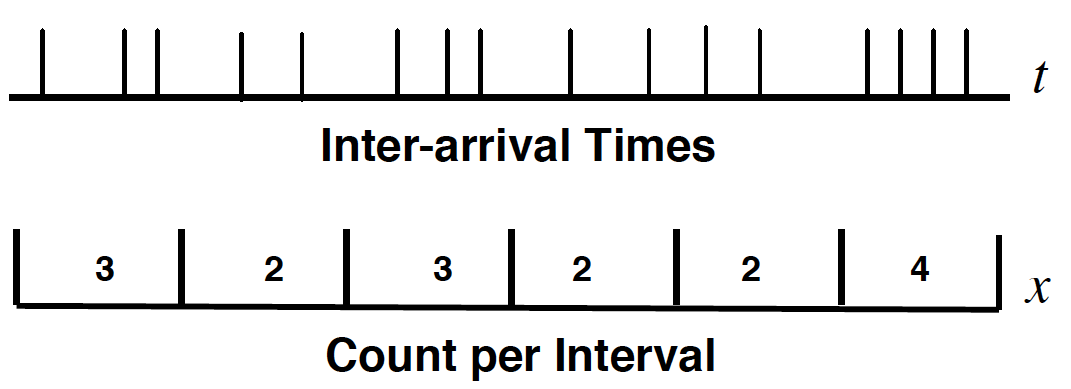} % for arXiv acceptance
\caption{Sample arrival pattern generated by a Poisson process} \label{fig:poisson-bins}
\end{figure}

We use the following properties of a \hyperlink{gloss:pproc}{Poisson process} 
%(see Glossary~\ref{sec:glosspproc}) 
in the subsequent sections.
\begin{enumerate}
\item \label{item:counts}
The {\em number} of  arrivals in each fixed interval of time is a random variable that is 
\hyperlink{gloss:pois}{Poisson distributed}.
%(see Glossary~\ref{sec:glosspois}).
This is shown as the {\em Count per interval} at the bottom of  Fig.~\ref{fig:poisson-bins}. 
The \hyperlink{gloss:pois}{Poisson distribution} is revealed by plotting the corresponding histogram. (See discussion in Appendix~\ref{sec:ruler})
\item   \label{item:exp}
The {\em periods} between arrivals has a random spacing that is \hyperlink{gloss:exp}{exponentially distributed}.
This is shown as the {\em Inter-arrival times} at the top of Fig.~\ref{fig:poisson-bins}.
The \hyperlink{gloss:exp}{exponential distribution} is revealed by plotting the corresponding histogram. (See Appendix~\ref{sec:ruler})
\item  \label{item:psuper}
Combining arrivals from $N$ \hyperlink{gloss:pproc}{Poisson processes} (e.g., web users), 
each having mean request rate 
$\lambda$, produces a single aggregated \hyperlink{gloss:pproc}{Poisson process} having a mean request rate 
$N \lambda$~\cite{KARL75,GUNT11}.
\item \label{item:usuper}
Combining $N$ non-Poisson processes asymptotically approaches an aggregated Poisson process provided $N$ is large and each $\lambda$ is very small.
This follows from the \href{https://en.wikipedia.org/wiki/Palm-Khintchine_theorem}{Palm-Kintchine theorem},  
which explains why so many natural and engineering phenomena can be well described by an aggregated 
\hyperlink{gloss:pproc}{Poisson process}~\cite{KLEI75}.  
\end{enumerate}

The clicks of a Geiger counter (\href{https://www.youtube.com/watch?v=upPiJ9vOYiY}{YouTube audio}) follow a \hyperlink{gloss:pproc}{Poisson process} due the random arrival of nuclear decay fragments at the detector. The clicks are analogous to requests arriving at the SUT.
An intuitive demonstration of the \hyperlink{gloss:pproc}{Poisson process}, based on random numbers generated in an Excel spreadsheet that are then mapped to points on a meter ruler, is provided in Appendix~\ref{sec:ruler} and discussed in~\cite{BRAD09}.

\subsection{Web testing principle B} \label{sec:princopsB}
The second principle that forms part of our methodology follows from Poisson property~\ref{item:exp} in the previous section, viz., the mean and standard deviation of an 
\hyperlink{gloss:exp}{exponential distribution} are equal. 
%(See Glossary~\ref{sec:glossexp}) 
More particularly, the ratio of the mean and standard deviation is called the 
\hyperlink{gloss:cov}{coefficient of variation}, 
or $CoV$, and therefore $CoV=1$ for an 
\hyperlink{gloss:exp}{exponential distribution}.  
%(see Glossary~\ref{sec:glosscov})
Since the measured times between web requests are expected to be 
\hyperlink{gloss:exp}{exponentially distributed}, 
statistical analysis  should produce a $CoV$ close to unit value.
This statistic forms the basis of \hyperlink{box:princB}{Principle B} for verifying that asynchronous web traffic 
generated according to \hyperlink{box:princA}{Principle A} is in fact produced during the load tests.

%%%%%%% Principle B  %%%%%%% 
\begin{mybox}
\begin{center} \hypertarget{box:princB}{\bf Principle B} \end{center}
Verify that the requests generated by applying \hyperlink{box:princA}{Principle A} (i.e. large $N$ with low rate $1/Z$)
closely approximate a Poisson process by measuring the coefficient of variation of the inter-arrival periods and demonstrating that $CoV \simeq 1$ within acceptable measurement error.
\end{mybox}
%%%%%%%%%%%%%%   

A synchronous or \hyperlink{gloss:closed}{closed} workload will exhibit a coefficient of variation $CoV < 1$ (see, e.g.,  Fig.~\ref{fig:jimfig10}) because the random variations or fluctuations in the Poisson arrival pattern are damped out by the negative-feedback loop discussed in Section~\ref{sec:methodology}. A \hyperlink{gloss:closed}{closed} workload is therefore {\em hypo}-exponential. 
According to \hyperlink{box:princA}{Principle A}, as the ratio $Q/Z \rightarrow 0$ and $\lambda \rightarrow N/Z$ in eqn.~\eqref{eqn:crate}, it follows that $CoV \rightarrow 1$ because the period between arrivals becomes exponentially distributed. This is how Principles A and B are connected.

Equation~\eqref{eqn:crate} {\em assumes} that the inter-arrival periods are exponentially distributed whereas \hyperlink{box:princB}{Principle B} {\em verifies} that assumption is not violated during the actual test.  
It is worth noting that our approach stands in contrast to \cite{SCHR06} where the focus is on the $CoV$ of the \hyperlink{gloss:stime}{service times} in the SUT; not the inter-arrival times. 
\hyperlink{gloss:stime}{Service demand} statistics are also generally more difficult to measure.

\subsection{Website case study} \label{sec:gov}
We now show how the principles developed in Sects.~\ref{sec:princopsA} and \ref{sec:princopsB} 
are applied to the actual load-testing of a real government website using 
\href{http://jmeter.apache.org}{JMeter} as the test harness.
The purpose of the government website is to allow state residents to obtain various kinds of government statistics. 
We abbreviate the website name to \gov\	for both simplicity and protection.  
Because  \gov\	was about to go live, all tests had to be performed in a single day, and the only test equipment available on such short notice was a relatively small load-driver capable of handling a JMeter thread group of  two hundred user threads, at most.
Despite these physical limitations in the test environment, 
the broader performance question of interest was, how many web users will the real \gov\ website be able to support? % embedded question form http://grammar.about.com/od/e/g/Embedded-Question-term.htm
We address that question in Sect.~\ref{sec:ausers}. 
Overall, this exercise became a balancing act between accurately emulating web traffic and  assessing website scalability beyond the data that could be captured within the limited testing time.
We address scalability projections in Sect.~\ref{sec:webscale}. 

\begin{table}[!ht]
    \caption{\gov\  test web pages and test web objects} \label{tab:webmap}
    \small 
    \begin{subtable}[t]{.5\linewidth} 
      \centering
        \caption{Web site test objects}  \label{tab:gov}
       	\begin{tabular}{l|l}
		\hline
		Object	& Definition\\
		\hline
		\verb|010_Home|	& Home page\\
		\verb|012_Home_jpg|	& Background image\\
		\verb|020_Dept|	& Department data\\
		\verb|022_Dept_jpg|	& Department image\\
		\verb|030_Demographics|	& Demographic data\\
		\verb|040_Statistics|	& Government user data\\
		\hline
		\end{tabular} 
    \end{subtable}%
    \begin{subtable}[t]{.5\linewidth}
      \centering
        \caption{Mapping between pages and  objects} \label{tab:map}
        \small 
       	\begin{tabular}{l|l}
		\hline
		Web Page	& Objects\\
		\hline
		Home	    & \verb|010_Home + 012_Home_jpg|\\
		Department	& \verb|020_Dept + 022_Dept_jpg|\\
		Demographics & \verb|030_Demographics|\\
		Statistics	& \verb|040_Statistics|\\
		\hline
		\end{tabular} 
    \end{subtable}%
\end{table}

In this scenario, the intent was for the \gov\ website to be reconfigured from standalone dedicated servers to a virtual-server environment. The load tests were based on a list of web-page objects specified by name and purpose in Table~\ref{tab:webmap}. 
%As noted in {\em Heterogeneous requests} of Sect.~\ref{sec:users}, 
Each \hyperlink{list:hetero}{heterogeneous} object is retrieved by an \get\ request. 
Based on the methodology described in Sections~\ref{sec:workloads} and~\ref{sec:methodology}, 
one goal of these load tests was  
to ensure that the web objects in Table~\ref{tab:gov} were retrieved in a manner consistent with real web users on the Internet.

\begin{figure}[!ht]
\centering
\includegraphics[scale = 0.5]{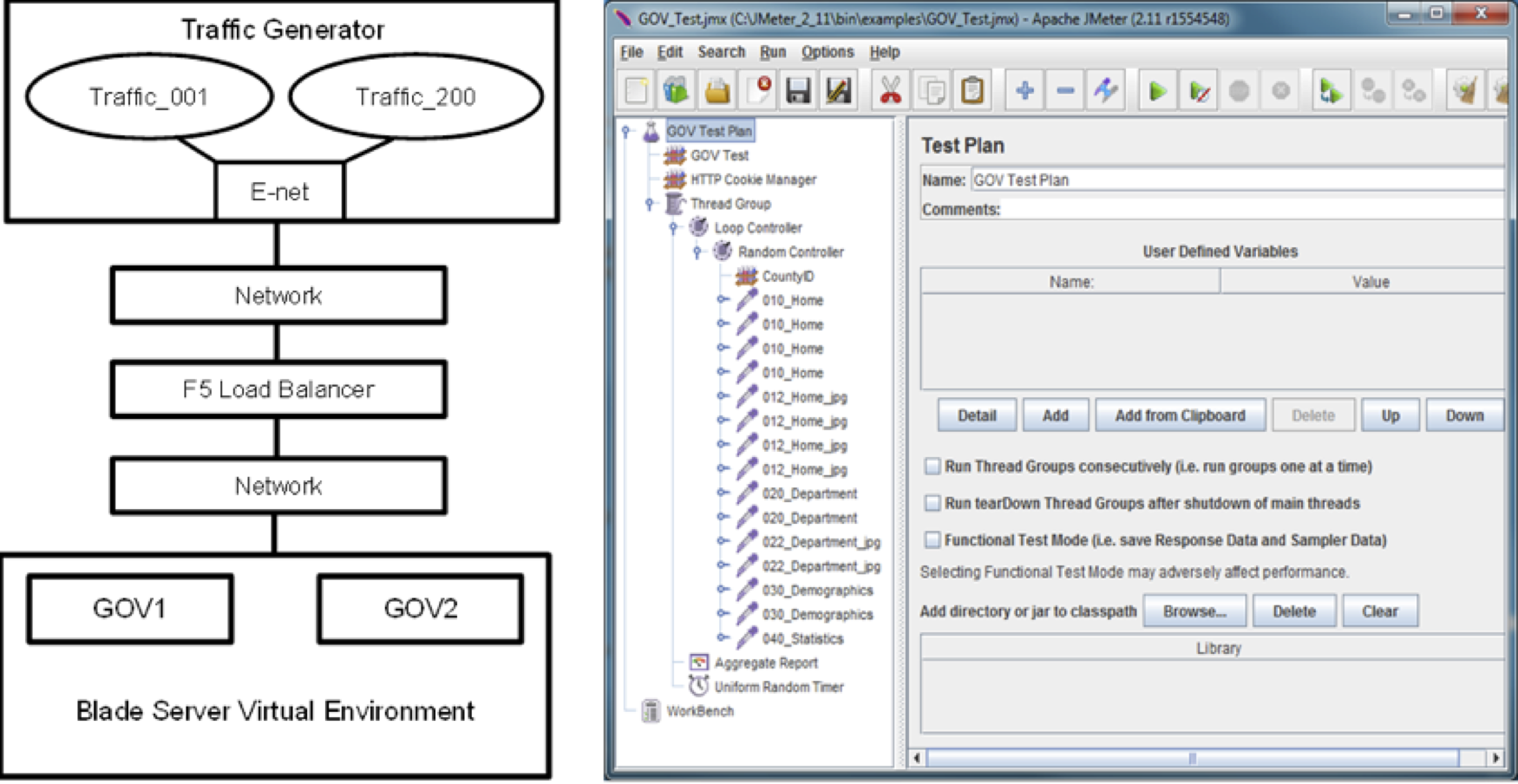} % for arXiv acceptance
\caption{\gov\ web site test configuration and JMeter script} \label{fig:jimfig7}
\end{figure}

The left side of Fig.~\ref{fig:jimfig7} is a topological view of the load testing environment showing load generator, network interfaces, 
\href{https://f5.com/glossary/load-balancer}{F5 load balancer}, and the virtualized blade server setup with \gov\ virtual servers GOV1 and GOV2. The right side of Fig.~\ref{fig:jimfig7} is the JMeter test script.

In addition to estimating how many web users can be supported, there was interest in 
determining the scalability of the virtual servers GOV1 and GOV2 
as well as evaluating how well the F5 balanced the load between them. 
The testing was performed under the following conditions:
\begin{itemize}
\item 
As the traffic-generator layout of Fig.~\ref{fig:jimfig7} implies, 
a maximum of $N=200$ JMeter threads (indicated by the \verb|Traffic_200| bubble)
was used due to driver-side operating system \hyperlink{list:driveros}{limits} 
%noted in {\em Client operating system} of Sect.~\ref{sec:users}.
\item
Web pages are accessed in random order with multiple instances of them implemented to bias the event count. For example, there are four \verb|010_Home page| object-instances compared with just one \verb|040_Statistics| object-instance.
\item
There is one {\em Uniform Random Timer} used to produce \hyperlink{gloss:ztime}{think times} spanning the entire thread group. 
A \hyperlink{gloss:uni}{uniform distribution} is a bounded distribution 
%(see Glossary~\ref{sec:glossuni}) 
so it terminates more gracefully at the end of each test run. Generator threads typically draw a 
\hyperlink{gloss:ztime}{think time} without checking to see if that time exceeds the wall-clock time of the test.
\item
Traffic is increased from test to test by reducing the mean \hyperlink{gloss:ztime}{think time} between tests.
\item
The seven 25-minute tests of increasing load are numbered 1800, 1830, 1900, 1930, 2000, 2030, 2100, 
according to the start time of a 24-hour clock. 
\item
Steady state measurements are ensured by excluding the first two and last three minutes of each test run.
This also mitigates any negative effects of the ramp-up and ramp-down phases in the data analysis.
\end{itemize}

\noindent
The traffic generator and the test script are structured so as to maintain a consistent traffic mix for scalability testing.
Objects are selected in random order to mimic the stateless HTTP protocol that supports an application that has no set sequence of page accesses.

Table~\ref{tab:jimfig8} summarizes the JMeter test parameters and inter-arrival statistics.
Rows 1--7 correspond to seven 25-minute test runs. 
Each test run employs the maximal number of JMeter generator threads. Hence, $N=200$ in our notation. 
Successive runs correspond to increasing the simulated web-user load which, in turn,
increases the steady-state throughput measured by JMeter, i.e., $X=\lambda$ objects per second in column 4. 
Since the maximal $N$ is constrained by the number of available \hyperlink{list:driveros}{driver threads},  traffic into the SUT is actually increased by successively decreasing $Z$ in column 3. 
As discussed in Sect.~\ref{sec:princopsA}, this is tantamount to increasing the emission rate $1/Z$ of each load generator.
Note that the number of \hyperlink{gloss:conc}{concurrent} requests in the SUT is very small 
(i.e., $Q \ll N$ in column 6) so, the key condition of \hyperlink{box:princA}{Principle A} is met. 
Also note that the total number of web objects in the test system, $N$  (obj) in column 7, can be estimated  as a decimal number using eqn.~\eqref{eqn:closedn} and is consistent with the integral number of actual JMeter threads, $N$ in column 2.
Even though the inter-arrival statistics appear to diverge from $CoV = 1$ at the highest traffic levels in this thread-limited environment, they are still within an acceptible range to conform to \hyperlink{box:princB}{Principle B} and prove that we are indeed simulating web-user traffic in accordance with 
\hyperlink{box:princA}{Principle A}.

\begin{table}[htp]
\begin{center} \footnotesize
\caption{\gov\ simulated web user loads} \label{tab:jimfig8}
\begin{tabular}{r rrr| rr | rr |rrr}
\hline										
& \multicolumn{3}{c|}{\bf JMeter Parameters}	& \multicolumn{2}{c|}{\bf JMeter Metrics} &  \multicolumn{2}{c|}{\bf Queue Metrics} & \multicolumn{3}{c}{\bf Inter-arrival Time Statistics}\\		
\hline
& Run	& $N$	& $Z$ (ms)	& $\lambda$ (obj/s)	& $R$ (ms) & $Q$ (obj)  & $N$ (obj) & Mean	(ms) & StdDev (ms) & $CoV$\\
\hline
1 & 1800	& 200	& 12500	& 15.91		&  53.17	& 0.85 & 199.72 & 62.81	& 61.75	& 0.98\\
\rowcolor{Goldenrod}
2 & 1830	& 200	& 6250	& 31.73		&  52.74	& 1.67 & 199.99 & 31.52	& 31.59	& 1.00\\
3 & 1900	& 200	& 4200	& 46.79		&  54.06	& 2.53 & 199.05 & 21.37	& 21.54	& 1.01\\
4 & 1930	& 200	& 3250	& 60.26		&  59.64	& 3.59 & 199.44 & 16.60	& 16.78	& 1.01\\
5 & 2000	& 200	& 2500	& 77.56		&  74.97	& 5.81 & 199.71 & 12.89	& 13.22	& 1.03\\
6 & 2030	& 200	& 1563	& 118.03	& 132.67    & 15.66 & 200.14 & 8.47	& 9.12	& 1.08\\
\rowcolor{Goldenrod}
7 & 2100	& 200	& 1000	& 159.16	& 253.38    & 40.33 & 199.49 & 6.28	& 7.37	& 1.17\\
\hline
\end{tabular}
\end{center}
\end{table}%

Clearly, the $CoV$ of the inter-arrival times is an important statistic   
needed to confirm that asynchronous web traffic is being produced by the load tests in Table~\ref{tab:jimfig8}. 
Unfortunately, that statistic is not provided by JMeter or any other load-test harnesses, to our knowledge.
Consequently, one of us (JFB) developed an analysis script to calculate the empirical $CoV$ by sorting logged JMeter requests into correct temporal  order and taking differences.

\subsection{Simulated web users} \label{sec:pool}
\begin{figure}[!h]
\centering
\includegraphics[scale = 0.4]{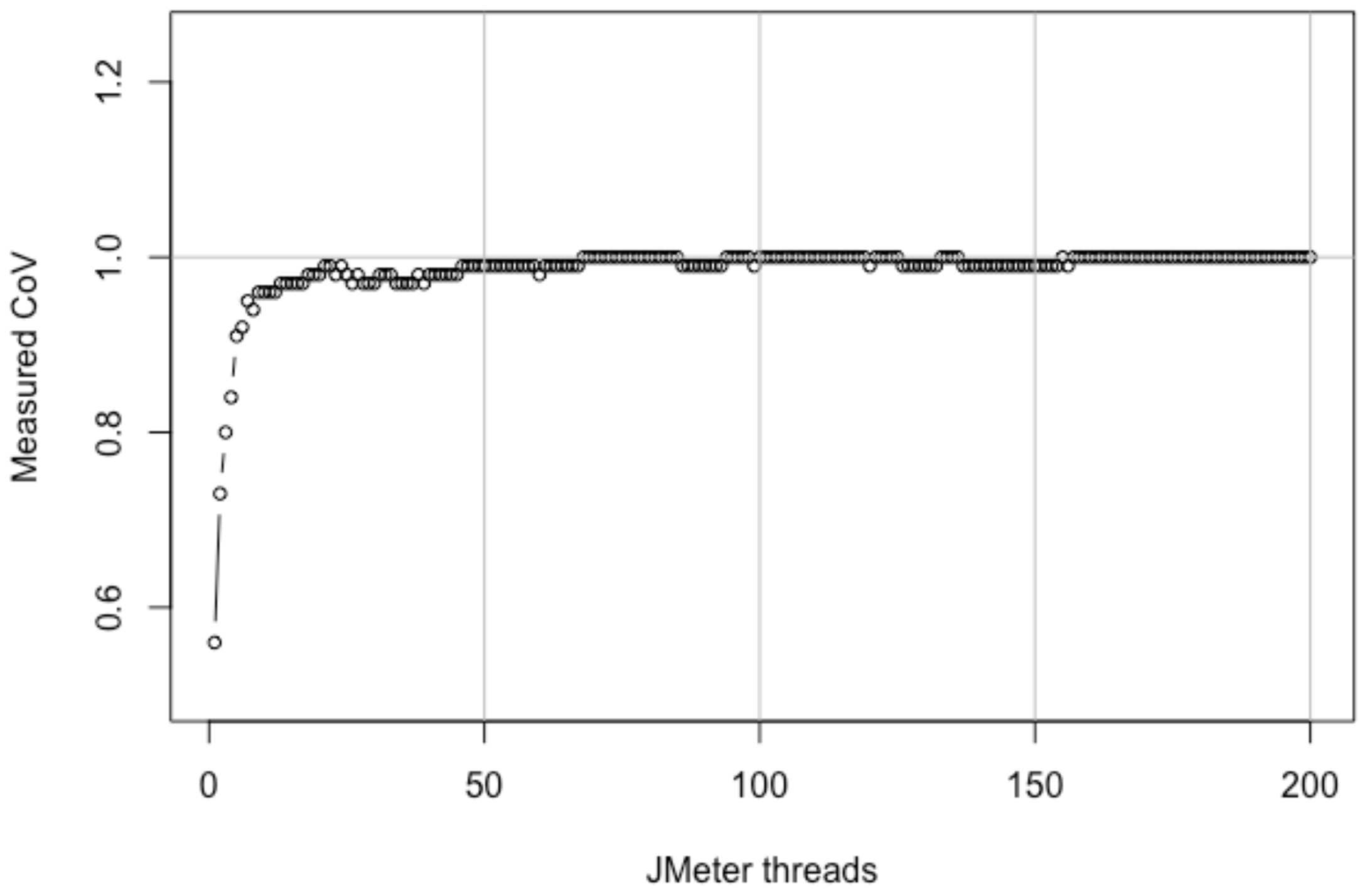} 
\caption{Inter-arrival time $CoV$ for JMeter threads up to $N = 200$ with $Z = 6.25$ seconds} \label{fig:jimfig10}
\end{figure}

The $CoV$ of inter-arrival periods in Table~\ref{tab:jimfig8} is based on a maximum pool of 
$N=200$ threads (column 2) allowed by the operating system that supports the load generators. 
Taking test-run number 1830 (row 2) as the example being closest to the required $CoV$,   
we can ask, how many threads are actually needed to reach unit value? 
Figure~\ref{fig:jimfig10} provides the answer. 

At around 50 threads, the monitored values start to settle down around $CoV=1$ all the way up to the maximum 200 threads. 
In other words, by applying \hyperlink{box:princA}{Principle A} of Sect.~\ref{sec:princopsA} in these tests, we see that the hypoexponential arrival pattern of the \hyperlink{gloss:closed}{closed} test environment is converted into the desired exponential arrival pattern using just 25\% of the total thread pool.
Note also that the data profile in Fig.~\ref{fig:jimfig10} is consistent with the rapid onset predicted by PDQ~\cite{GUNT15} in Fig.~\ref{fig:NZlimit}. 

Poisson property~\ref{item:psuper}, in Sect.~\ref{sec:poissonprops}, supports the derivation  of 
eqn.~\eqref{eqn:zsum} which assumed $N$ \hyperlink{gloss:indpt}{independent} virtual users,  
%(see Glossary~\ref{sec:glossindpt}), 
each issuing requests with a mean rate  $1/Z$, can be added together at the SUT and treated as though they are requests coming from a single user at a mean rate $N/Z$. This Poisson property is guaranteed to be satisfied when the think-time periods are exponentially distributed with mean time $Z$ (i.e., Poisson property~\ref{item:exp}). 

However, JMeter, like many load-test tools, does not offer an exponential-timer option~\cite{GUNT10b,BRAD12}. 
The test configuration specified in the previous section used uniformly distributed think-time periods. 
In other words, the actual load generators appear to be in violation of our assumed Poisson property~\ref{item:psuper}.
Poisson property~\ref{item:usuper} provides the escape hatch.  
As long as $1/Z$ corresponds to a very low rate of requests and $N$ is relatively large, an aggregate Poisson process will still be well approximated~\cite{KARL75,ALBI82}. 
As seen in Fig.~\ref{fig:jimfig7}, the JMeter Uniform Random Timer was used in this case study and 
Fig.~\ref{fig:jimfig10} confirms that Poisson property~\ref{item:usuper} applies to a uniform think-time distribution.

\subsection{Internet web users} \label{sec:ausers}
Since the main question of interest to management was the predicted number of real web users that  \gov\ could be expected to support, we leveraged the results of Sect.~\ref{sec:pool} to move from simulated web users in the test environment to real web users on the Internet.
To avoid any confusion with our earlier notation, we denote real web users by $N^\prime$ and retain $N$ for simulated web users. Since we cannot measure $N^\prime$ directly 
(cf. the clouds in Figs.~\ref{fig:smiles} and~\ref{fig:mm1}), we calculate it as a statistical quantity using 
\begin{equation}
N^\prime = \lambda (R + Z^\prime)  \label{eqn:jn}
\end{equation}
which is the analog of eqn.~\eqref{eqn:closedn}.
We already know $\lambda$ and $R$ from the simulated web-user measurements in Table~\ref{tab:jimfig8}. 
The \hyperlink{gloss:ztime}{think time} $Z^\prime$ is chosen to reflect real-world delays, e.g., Internet latencies and the time taken by human users when perusing actual web pages.
Since $N^\prime$ is a statistical mean, it appears as a decimal number 
(rather than an integer that would normally be used to specify the number of load generators)  
in Table~\ref{tab:jimfig11}. 

\begin{table}[h!]
\begin{center} \footnotesize
\caption{\gov\ estimated Internet web users} \label{tab:jimfig11}
\begin{tabular}{r rrrr||rr|rrrrr}
\hline
 & \multicolumn{4}{c||}{ } & & & \multicolumn{5}{c}{{\bf Nominal Internet Think Time, $Z^\prime$} (ms)} \\ \cline{8-12}	
%\hline			
 & \multicolumn{4}{c||}{\bf JMeter Measurements} & \multicolumn{2}{c|}{\bf Conversion} &	10,000	& 20,000 & 30,000 & 40,000	&  50,000\\
\hline
 & Run & $\lambda$ (obj/s) & R (ms) & Q (obj) & $\lambda$ (pgs/s) & Q (pgs) & \multicolumn{5}{c}{ {\bf Calculated Internet Web Users, $N^\prime$}}	\\
\hline		
1  & 1800 &  15.91 &  53.17 &  0.85 &  9.5460 &  0.51 &  95.97 & 	191.43 & 	286.89 & 	382.35 & 	477.81\\
2  & 1830 &  31.73 &  52.74 &  1.67 & 19.0380 &  1.00 & 191.38 & 	381.76 & 	572.14 & 	762.52 & 	952.90\\
3  & 1900 &  46.79 &  54.06 &  2.53 & 28.0740 &  1.52 & 282.26 & 	563.00 & 	843.74 & 	1124.48 & 	1405.22\\
\rowcolor{Goldenrod}
4  & 1930 &  60.26 &  59.64 &  3.59 & 36.1560 &  2.16 & 363.72 & 	725.28 & 	1086.84 & 	1448.40	& 1809.96\\
5  & 2000 &  77.56 &  74.97 &  5.81 & 46.5360 &  3.49 & 468.85 & 	934.21 & 	1399.57 & 	1864.93 & 	2330.29\\
6  & 2030 & 118.03 & 132.67 & 15.66 & 70.8180 &  9.40 & 717.58 & 	1425.76 & 	2133.94 & 	2842.12 & 	3550.30\\
\rowcolor{Goldenrod}
7  & 2100 & 159.16 & 253.38 & 40.33 & 95.4960 & 24.20 & 979.16 & 	1934.12 & 	2889.08 & 	3844.04 & 	4799.00\\
\hline
\end{tabular}
\end{center}
\end{table}%

Rows 1--7 in Table~\ref{tab:jimfig11} correspond to the same seven tests shown in Table~\ref{tab:jimfig8}. 
Some web objects are requested as an aggregate by the browser when a real user performs a web-page query. 
See {\em Heterogeneous requests} of Sect.~\ref{sec:users}. 
Table~\ref{tab:map} lists the mapping of web pages to web objects for the \gov\ web site and shows that the Home web page and the Department web page each contain two web objects.
Because $\lambda ~\text{(obj/s)}$, as measured by JMeter,  
is actually the \hyperlink{gloss:arate}{arrival rate} of web objects (see column 2 of Table~\ref{tab:jimfig11}), and not web pages, the web-page per web-object ratio needs to be determined. The JMeter script in Fig.~\ref{fig:jimfig7} contains 15 web objects. 
When these objects are consolidated, 
the result is $4~Home~pages + 2~Department~pages + 2~Demographics~pages + 1~Statistics~pages = 9$ web pages. The measured \hyperlink{gloss:arate}{arrival rate} $\lambda~\text{(obj/s)}$ is multiplied by this 9:15 ratio to convert it into the necessary web-page \hyperlink{gloss:arate}{arrival rate}, $\lambda~\text{(pgs/s)}$.
This pro-rated \hyperlink{gloss:arate}{arrival rate} is used to calculate the average of number of Internet web users, $N^\prime$, in  Table~\ref{tab:jimfig11}. 

\begin{exam}[Test Run 2100]
Consider the load test data in row 7 of Table~\ref{tab:jimfig11}. 
The \hyperlink{gloss:arate}{arrival rate} \mbox{$\lambda = 159.16$} web-objects per second in column 2 comes from  row 7 of 
Table~\ref{tab:jimfig8}. 
Applying the 9:15 conversion ratio yields $9 \times 159.16 / 15 =95.50$ web-pages per second as the request rate in the 5th column.
The measured \hyperlink{gloss:rtime}{response time} is 253.38 milliseconds (column 3). 
Substituting an Internet user \hyperlink{gloss:ztime}{think time} of $Z^\prime = 30,000$ milliseconds (from column 9) into eqn.~\eqref{eqn:jn}, we find $N^\prime = 95.4960 \times (0.25338 + 30) = 2889.08$ mean Internet web users. 
\end{exam}

In Sect.~\ref{sec:princopsA} it was pointed out that $Q \ll N$ is a condition for eqn.~\eqref{eqn:crate} to be a valid approximation to web traffic. We see that condition reflected as $Q < N^\prime$ in Table~\ref{tab:jimfig11}. Similarly, by virtue of Little's law, $R < Z^\prime$ in eqn.~\eqref{eqn:jn}.
We can also verify that our estimates of $N^\prime$ and $Z^\prime$ scale in conformance with \hyperlink{box:princA}{Principle A}. 

\begin{exam}[Scaled-Z arrival rate]
Consider test run 1930 (row 4 of Table~\ref{tab:jimfig11}) for Internet user \hyperlink{gloss:ztime}{think time}s $Z^\prime = 10,20, \dots, 50$ seconds (top of columns 7--11). 
Applying eqn.~\eqref{eqn:crate}, the expected \hyperlink{gloss:arate}{arrival rate} into \gov\ is
\begin{align*}
\lambda_{10} &= \dfrac{363.72}{10}  \quad - \;\dfrac{2.16}{10} = 36.16\\
\lambda_{20} &= \dfrac{725.28}{20}  \quad - \;\dfrac{2.16}{20} = 36.16\\
\lambda_{30} &= \dfrac{1086.84}{30} \;\; -  \;\dfrac{2.16}{30} = 36.16\\
\lambda_{40} &= \dfrac{1448.40}{40} \;\: -  \; \dfrac{2.16}{40} = 36.16\\
\lambda_{50} &= \dfrac{1809.96}{50} \;\: -  \; \dfrac{2.16}{50} = 36.16
\end{align*}
all of which are the same constant mean rate and in agreement with the value shown in column 5.
\end{exam}

\noindent
Under different circumstances and test requirements, load tools like those identified in 
Sect.~\ref{sec:tools} could be used to validate these estimates. In the meantime, our methodology provides 
a low-cost alternative.

\subsection{Website scalability}  \label{sec:webscale}
\begin{figure}[!ht]
\centering
\includegraphics[scale = 0.525]{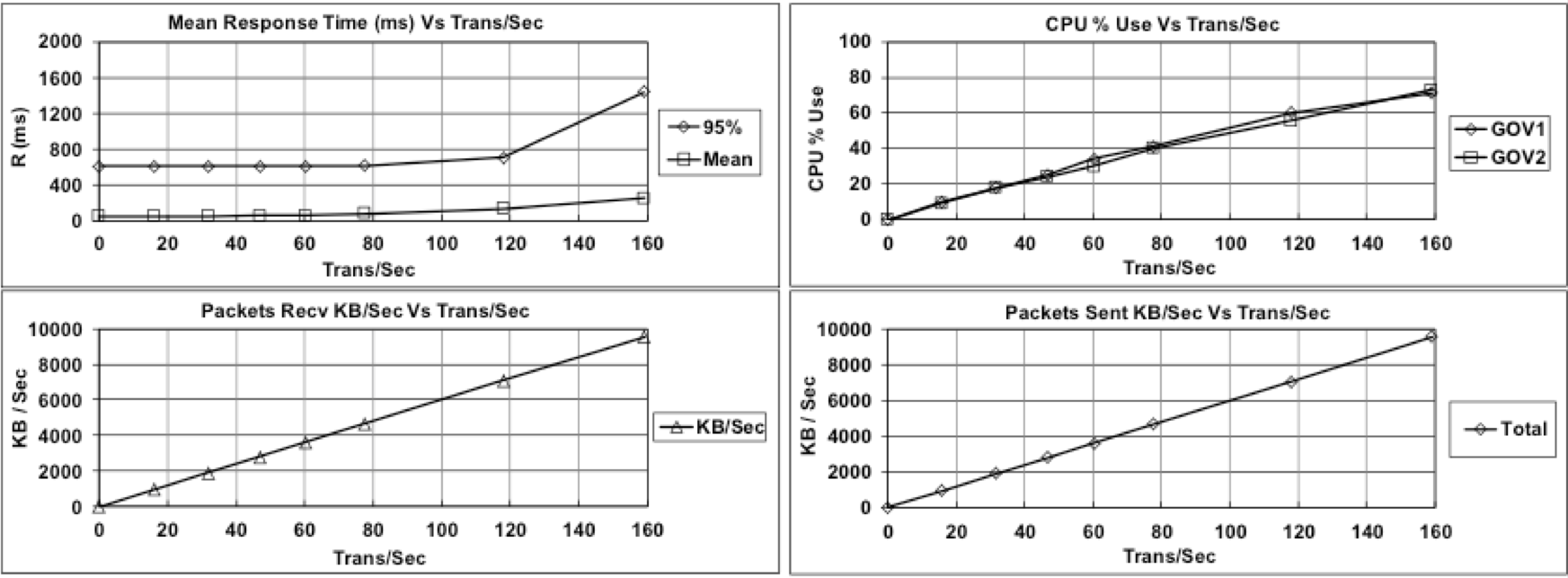} % for arXiv acceptance
\caption{\gov\  load balancing and scalability data} \label{fig:jimfig12}
\end{figure}

Figure~\ref{fig:jimfig12} indicates that the \gov\ virtual environment is load balanced and capable of handling 159.16 web \get\ or \post\ requests per second with all CPUs running around 70\% busy.
Viewing the 200 JMeter threads as asynchronous transaction generators, instead of canonical virtual users, 
shifts the focus of SUT load balance and scalability analysis from users to the traffic they produce. 
Hence, the x-axes of the plots are labelled by transactions per second (TPS), not users. 
This relabeling is also consistent with the typical units of rate that are used in the context of \hyperlink{gloss:open}{open} queueing systems~\cite{GUNT11}.

The plot in the upper-left quadrant of Fig.~\ref{fig:jimfig12} contains the mean and 95th percentile of 
\hyperlink{gloss:rtime}{response time} (as measured by JMeter) plotted as a function of TPS for the seven test runs performed in Sect.~\ref{sec:pool}. 
Both response-time curves are relatively flat until test run 2100, when the hockey-stick handle corresponding to 
Fig.~\ref{fig:rt-asymptote}, begins to appear at a mean $R$ of 253 milliseconds and an  
$R_{95}$ of 1440 milliseconds. Although these levels are larger than baseline, they were considered to be acceptable for this particular application.

The plot in the lower-left quadrant of Fig.~\ref{fig:jimfig12} 
shows network packets received in KB/s, as measured by JMeter. 
These values are higher than send-traffic volumes because the data size of the web server responses are almost always larger than \get\ or \post\ requests. 
The plot shows that the approximately 10,000 KB/s network link (i.e., 100 Mbps packet bandwidth) is nearly saturated as test run 2100 reaches a throughput of 9,552 KB/s.

The plot in the upper-right quadrant of Fig.~\ref{fig:jimfig12} 
shows CPU utilization as a function of TPS for each of the two virtual servers. The two plots 
essentially lie on top of each other for all seven tests, indicating a load balanced environment. 
Moreover, both plots are linear through the 71\% CPU busy, achieved at test run 2100, thus indicating excellent scalability up to that utilization level.

The plot in the lower-right quadrant of Fig.~\ref{fig:jimfig12} 
represents combined SUT network packets sent in KB/s. 
The virtualized servers, GOV1 and GOV2 (see Fig.~\ref{fig:jimfig7}), sent network packets at an aggregate rate of 9,538 KB/sec in test run 2100 on two 10,000 KB/s  (or 100 Mbps) network links. 
These data provide a statistical cross-check that packet traffic leaving the SUT is the same as the 9552 KB/s received by JMeter.

\section{Conclusion} \label{sec:summary}
Interactive computing has evolved over the past fifty years from a relatively small number of centralized users issuing commands into a monolithic application via their fixed terminals, 
to a much larger number of remote users issuing requests into web services via a plethora of personal computing devices.
Conventional load-test tools are designed to mimic the former rather than the latter and  
that creates a problem for performance engineering when it comes to 
accurately load-testing web applications. 
In Sections~\ref{sec:methodology} and~\ref{sec:application}, we presented a methodology to address that problem.

Our methodology is based on two fundamental principles. 
\hyperlink{box:princA}{Principle A}, in Sect.~\ref{sec:princopsA}, tells us how to scale the parameters $N$ and $Z$ so that the test environment is optimized for producing requests that mimic the Poisson process associated with web-user workloads.
\hyperlink{box:princB}{Principle B}, in Sect.~\ref{sec:princopsB}, tells us how to confirm that the test parameter settings did actually mimic a Poisson process during the course of each test by analyzing the $CoV$ of the measured inter-arrival times.
In practice, the request rate $\lambda$ into the SUT can be increased in any of three ways:  
\begin{inlinelist}
  \item increase $N$ while holding $Z$ fixed (see Table~\ref{tab:zconst}), 
  \item decrease $Z$ while holding $N$ fixed (see Table~\ref{tab:jimfig8}), 
  \item hold the ratio $N/Z$ fixed while increasing $N$ (see Table~\ref{tab:zscaled}).
\end{inlinelist}
Of these, the third option offers the royal road to emulating web-user traffic.

The development of \hyperlink{box:princB}{Principle B} also extends the work of ~\cite{ALBI82}, who analyzed a $\sum_N GI_N$/M/1 queue 
using an event-based simulator with up to 1024 aggregated iid arrival processes. 
The aggregated non-Poisson arrivals potentially represent a more realistic queueing model of a web server than the M/M/1 queue discussed in Sect.~\ref{sec:open} or the M/M/1/N/N queue discussed in Sect.~\ref{sec:closed}.  
However, the simulation suite did not include uniformly distributed inter-arrival times. 
Consistent with the Palm-Kintchine theorem in Sect.~\ref{sec:poissonprops}, $\sum_N GI_N$/M/1 diverges from M/M/1 for server utilizations greater than 50\% but, as we have shown using JMeter as a more realistic workload simulator in Sect.~\ref{sec:pool}, the slowest divergence occurs for aggregated non-Poisson arrivals that are uniformly distributed with a low $1/Z$ rate.

Our results offer a different perspective on the choice of queueing system for modeling asynchronous arrivals. To model M/M/1 exactly, in a load-test environment, would require significant modification to the generator scripts so that they issue requests randomly in time and are independent of any outstanding requests in the SUT.
However, as we showed in Sect.~\ref{sec:gov}, real Internet-based web users do actually submit synchronous requests into the website, but the synchronization is relatively weak: precisely as \hyperlink{box:princA}{Principle A} prescribes.
Somewhat surprisingly then, M/M/1/N/N with low $1/Z$ rate is a better queue-theoretic approximation to real web users than an explicit M/M/1 model, and $R < Z$ is consistent with HTTP being a stateless protocol.
On the other hand, one might choose to use the less realistic M/M/1 approximation simply because  
the actual values of $N$ and $Z$ are not known, as in Sect.~\ref{sec:ausers}.

We applied our methodology to load testing a major web site in Section~\ref{sec:application}. 
In practice, as discussed in Sect.~\ref{sec:ausers}, 
an estimation of actual $N^\prime$ Internet users---rather than the number of test generators, $N$---may need to be reported as the performance metric of interest. 
In Sections~\ref{sec:users},~\ref{sec:closed}, and~\ref{sec:ausers}, we attempted to clarify the meaning of user-\hyperlink{gloss:conc}{concurrency} in this context. % 2.3, 3.2 and 4.3
As a further aid to performance engineers, the statistical analysis tools we developed 
are available \href{https://github.com/DrQz/web-generator-toolkit}{online}.

%====================================================
\section{Acknowledgements}
%\section{Trademarks}
%%% Acknowledgements
% go here
We thank G. Beretta, S. Keydana, S. Parvu and CMG referees for their comments on earlier drafts. 
All brands and products referenced in this document are acknowledged to be the trademarks or registered trademarks of their respective holders.

%====================================================

\appendix
\appendixpage
\addappheadtotoc

\section{Meter Ruler Illustration} \label{sec:ruler}
\begin{figure}[!ht]
\centering
\includegraphics[scale = 1.0]{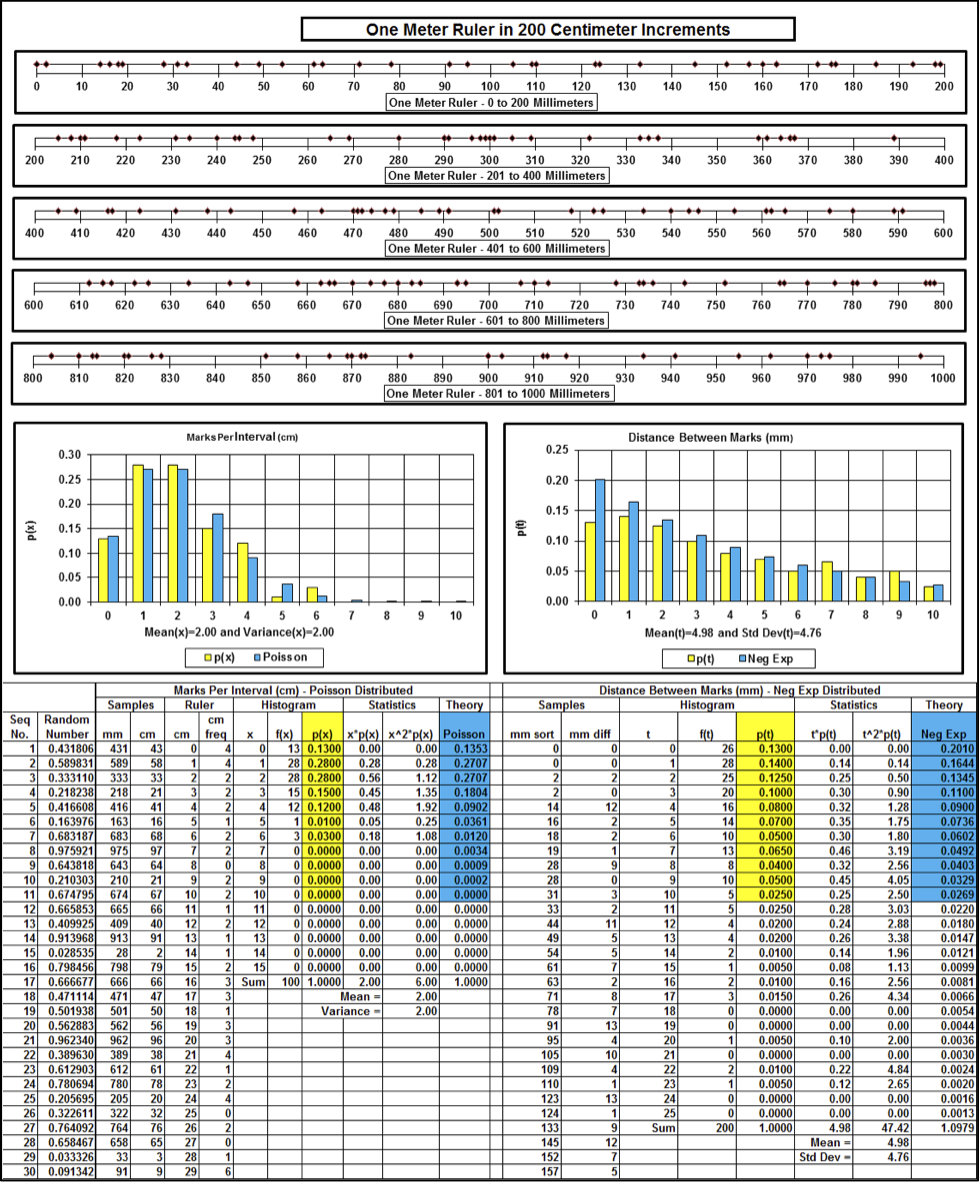} 
\caption{One meter ruler spreadsheet illustration} \label{fig:jimfigAppdx}
\end{figure}

Figure~\ref{fig:jimfigAppdx} is a one meter ruler spreadsheet illustration of random arrivals where 200 samples are drawn using the Excel rand() function and mapped to millimeter values shown as marks on the ruler. The spreadsheet lists the first 30 out of the 200 samples drawn as well as contains graphs and statistics for ``Marks per Interval (cm)'' and ``Distance Between Marks (mm).'' Both the ``Marks Per Interval'' and ``Distance Between Marks'' bar charts are accompanied by their theoretical distributions, viz., 
the \hyperlink{gloss:poiss}{Poisson} and \hyperlink{gloss:exp}{negative-exponential}. 

The spreadsheet columns (left to right) contain random numbers along with corresponding ruler mark values, mark per centimeter frequency statistics and associated histogram ordinates. This information is followed by histogram statistics and \hyperlink{gloss:poiss}{Poisson distribution} ordinate values. The distance between mark ordinates, histogram, histogram statistics, and 
\hyperlink{gloss:exp}{exponential distribution} ordinates are developed by first sorting the mark values in ascending order and recording the difference between adjacent marks.

An inspection of the one meter ruler reveals 13 one centimeter intervals with 0 marks, $f(x) = 13$, and an analysis of the distance between marks section of the full spreadsheet yields 26 places with 0 millimeters between marks, $f(t) = 26$.

The `Marks Per Interval' bar chart on the left below the one meter ruler contains frequency data in yellow and theoretical distribution values in blue yielding a mean = 2.00 cm and Var = 2.00 cm. 
The {\em Distance Between Marks} bar chart on the right shows a mean = 4.98 mm and an SD = 4.76 mm. These graphs indicate the mean and variance are equal for "Marks Per Interval" and the mean and standard deviation are statistically the same for "Distance Between Marks". Both are properties of a 
\hyperlink{gloss:pproc}{Poisson process} or a random arrivals environment. 
%defined in Appendix~\ref{sec:glosspproc}.

\section{Glossary} \label{sec:glossary}

\indent \indent % need 2 of these for 1st para
\hypertarget{gloss:arate}{{\bf Arrival rate:}}
%\subsection{Arrival rate} 
The rate $\lambda = A/T$ where $A$ is the {\bf count} of arrivals and $T$ is the measurement period. The same rate is used to parameterize a \hyperlink{gloss:pproc}{Poisson process}.

\begin{figure}[!h]
\centering
\includegraphics[scale = 0.5]{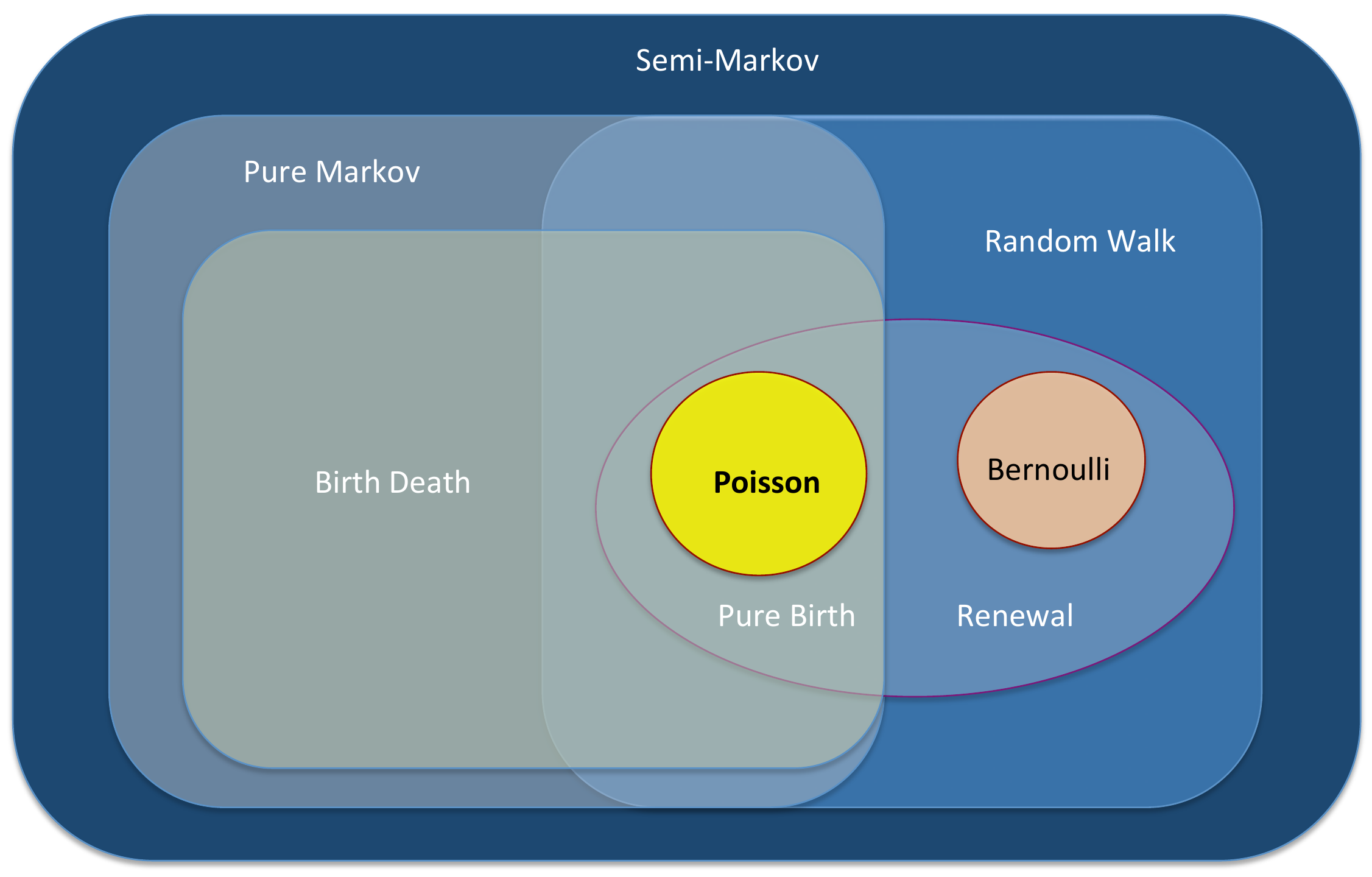} 
\caption{Venn diagram of the relationship between some stochastic processes} \label{fig:stochprocs} 
\end{figure}

%\noindent
\hypertarget{gloss:bern}{{\bf Bernoulli process:}}
A random sequence of events where the outcome is either a 
``success'' or a ``failure'', e.g., a head or a tail in the toss of a fair coin. 
Each coin toss---often called a {\em Bernoulli trial}---is assumed to be a 
{\em statistically \hyperlink{gloss:indpt}{independent}} 
event in that the resultant head or tail of one toss has no influence or bearing on the outcome of the next toss. 
The probability of getting a run of $k$ heads, for example, is given by the Binomial distribution.
Real coins are not fair and real coin tosses are not entirely statistically independent. 
A Bernoulli process can also be characterized as a sequence of iid 
\hyperlink{gloss:geo}{geometrically distributed} inter-arrival intervals. (See Fig.~\ref{fig:stochprocs})
% and Glossary~\ref{sec:glossindpt}.
Other important stochastic processes are shown in Fig.~\ref{fig:stochprocs}.

%\noindent
\hypertarget{gloss:bin}{{\bf Binomial distribution:}}
The distribution associated with discrete events that have a binary outcome with probability of success, $p$ and probability of failure, $q=1-p$. For a fair coin toss (a \hyperlink{gloss:bern}{Bernoulli process:}), $p=q=1/2$.
\begin{enumerate}
\item Discrete random variable $k=0,1,2,\ldots$
\item PMF $f(k) = \binom{n}{k} \, p^kq^{n-k}$ where $q=(1-p)$
\item CDF $F(k)= \sum_{i=0}^k \binom{n}{i} p^i q^{n-i}$
\item Mean $E(k) = np$
\item Variance $Var(k) = npq$
\item Standard deviation  $SD = \sqrt{npq}$
\item Coefficient of variation $CoV = \dfrac{q}{\sqrt{npq}}$
\end{enumerate}

%\noindent
\hypertarget{gloss:closed}{{\bf Closed system:}}
A queueing system with a finite number of requesters. 
No additional requests are allowed to enter from outside the system. (cf. \hyperlink{gloss:open}{open} queues)
Provides an elementary performance model of a load-test system.
Since no more than a single request can be outstanding, 
the limited number of requests come from the finite number of load generators. 
This means a closed system is actually comprised of two queues: generators (with no waiting line) and 
a queueing facility (the system under test). See Figs.~\ref{fig:frowns} and~\ref{fig:mm1n}.

%\noindent
\hypertarget{gloss:cov}{{\bf Coefficient of variation:}}
Denoted by $CoV$, it is the ratio of the standard deviation and the mean. 
For an analytic distribution $CoV = \sigma / \mu$, where $\sigma$ is the standard deviation and  
$\mu$ is the mean of the distribution. For measurements, of the type performed in this paper, 
$CoV = SD / M$, where $SD$ is the sample standard deviation and $M$ the sample mean.

%\noindent
\hypertarget{gloss:conc}{{\bf Concurrency level:}}
Denoted by $Q$, it is the combined number of requests waiting and the number of requests in service.

%\noindent
\hypertarget{gloss:exp}{{\bf Exponential distribution:}}
The distribution associated with the inter-arrival times from a \hyperlink{gloss:pproc}{Poisson process} as well as the service periods of an M/M/m queue. It has the following properties:
\begin{enumerate}
\item Continuous random variable $x$
\item PDF $f(x) = \lambda e^{-\lambda x}$
\item CDF $F(x) = 1 - e^{-\lambda x}$
\item Mean $E(x) = 1/\lambda$
\item Variance $Var(x) = 1/\lambda^2$
\item Standard deviation $SD = 1/\lambda$
\item Coefficient of variation $CoV = 1$
\end{enumerate}

%\noindent
\hypertarget{gloss:geo}{{\bf Geometric distribution:}}
The probability distribution of the number of 
\hyperlink{gloss:bern}{Bernoulli} trials needed to obtain a single success. 
\begin{enumerate}
\item Discrete random variable $k=0,1,2,\ldots$ with probability of success $p \in (0,1]$
\item PMF $f(k) = p \, q^{k-1}$ where $q=(1-p)$
\item CDF $F(k) = 1-q^{k}$
\item Mean $E(k) = 1/p$
\item Variance $Var(k) = q / p^{2}$
\item Standard deviation  $SD = \sqrt{q} / p$
\item Coefficient of variation $CoV = \sqrt{q}$
\end{enumerate}

%\noindent
\hypertarget{gloss:indpt}{{\bf Independence (statistical):}}
The innocent looking word ``independent'' appears in different guises throughout the literature on probability and statistics (including this paper) and has several special meanings.
Broadly speaking, it refers to the assumption that there are no correlations between values of the random variable of interest, e.g., arrival events of a \hyperlink{gloss:pproc}{Poisson process}. 
The more technical phrase is: random variables are 
{\em iid} or independently and identically distributed. 

For example, the number of arrival events contained in each uniformly spaced bin of Fig.~\ref{fig:poisson-bins} is assumed to be {\em iid} according to a 
\hyperlink{gloss:poiss}{Poisson distribution}. Note, it does not mean they are identically distributed, otherwise the numbers in each bin would all be the same. Rather, it means they are distributed with the same (identical) probability. 

The term {\em independent} is often used as a proxy for the word {\em random} (which is a much deeper subject). 
Another way to think of statistical independence for a \hyperlink{gloss:pproc}{Poisson process} is to imagine that the width of the uniform bins in Fig.~\ref{fig:poisson-bins} is made smaller and smaller until any bin can contain just a single arrival, at most. Many of these small bins will be empty. If a bin contains an arrival, that is like a \hyperlink{gloss:bern}{Bernoulli} success. 
The sequence of randomly occupied time-bins is like the random heads in a sequence of
\hyperlink{gloss:bern}{Bernoulli} coin tosses, each of which is assumed to be independent.

%\noindent
\hypertarget{gloss:open}{{\bf Open system:}}
A queueing system where an unlimited number of requests are allowed to enter from outside the system. Provides an elementary performance model of a web-based HTTP server where ``outside'' refers to the Internet. See Figs.~\ref{fig:smiles} and~\ref{fig:mm1}. Example open queueing models discussed in this paper include 
M/M/1, M/G/1~\cite{KLEI75,GUNT11} and $\sum_N GI_N$/M/1~\cite{ALBI82}.

%\noindent
\hypertarget{gloss:pois}{{\bf Poisson distribution:}}
The distribution associated with the number of arrivals from a \hyperlink{gloss:pproc}{Poisson process} 
as well as the service periods of an M/M/m queue. It has the following properties:
\begin{enumerate}
\item Discrete random variable $k=0,1,2,\ldots$
\item PMF $f(k,\alpha) = \dfrac{\alpha^k}{k!} \, e^{-\alpha}$
\item CDF $F(x,\alpha) =  \displaystyle \sum_{i=0}^{k} \, \dfrac{\alpha^i}{i!} \, e^{-\alpha}$
\item Mean $E(k) = \alpha$
\item Variance $Var(k) = \alpha$
\item Standard deviation $SD = \sqrt{\alpha}$
\item Coefficient of variation $CoV = 1/\sqrt{\alpha}$
\end{enumerate}
A simple mnemonic for the Poisson distribution is given in~\cite{GUNT14a}.

%\noindent
\hypertarget{gloss:pproc}{{\bf Poisson process:}}
A stochastic process is a dynamical map in either space or time that is generated by the values of a random variable. Some important stochastic processes are shown in Fig.~\ref{fig:stochprocs}. 
As employed in this paper, a Poisson process refers to the generation of random events that are counted in 
continuous time, $N(t)$. The counts are commonly referred to as {\em arrivals} and since they  
accumulate, $N(t)$ is an increasing stochastic function of time.
More formally the following three conditions need to be met:
\begin{enumerate}
\item $N(0) = 0$.
\item $N(t)$ has \hyperlink{gloss:indpt}{independent} increments.
\item The number of arrivals in any time interval has a \hyperlink{gloss:pois}{Poisson distribution}. 
\end{enumerate}
The number of arrivals in any time interval depends only on the length of the chosen interval, not on its location on the continuous time-line.

The relationship to the \hyperlink{gloss:exp}{exponential distribution} 
%in Glossary~\ref{sec:glossexp} 
can be seen as follows.
A Poisson process creates arrivals at random time-intervals $A_1,A_2, \ldots$ seen as the vertical lines in the top part of Fig.~\ref{fig:poisson-bins}.
We want to know the distribution  of these times $A_i$.
If there are no arrivals during the interval $[0,t]$,  
then the first arrival must occurs {\bf after} time $t$, i.e., 
\begin{equation*}
\Pr(A_1 > t) \equiv \Pr(N(t) = 0)
\end{equation*}
otherwise, it would not be the first arrival. 
From the \hyperlink{gloss:pois}{Poisson distribution} 
with $\alpha = \lambda t$ and $k=0$ (i.e., no event), this probability is given by the Poisson CDF:
\begin{equation*}
\Pr(A_1 > t) = \dfrac{(\lambda t)^0}{0!} \, e^{-\lambda t}  = e^{-\lambda t} 
\end{equation*}
The complementary probability, $\Pr(A_1 \leq t)$, that an arrival {\em did} occur in $[0,t]$, is then given by
\begin{equation*}	
1 - P(A_1 > t) = 1 - e^{-\lambda t}
\end{equation*}	
which is just the \hyperlink{gloss:exp}{exponential} CDF 
%in Glossary~\ref{sec:glossexp} 
with $x$ replaced by $t$.
Other key features of a Poisson process are discussed in Section~\ref{sec:poissonprops}.

%\noindent
\hypertarget{gloss:reztime}{{\bf Residence time:}}
Denoted by $R = W + S$ is the total average time spent in a single queue. 
(cf. \hyperlink{gloss:rtime}{response time})

%\noindent
\hypertarget{gloss:rtime}{{\bf Response time:}}
The sum of the \hyperlink{gloss:reztime}{residence times}, $\sum_k R_k$, belonging to a sequence of $k$ queueing centers. The response time is identical to the residence time for a single queueing center (i.e., $k=1$). 

%\noindent
\hypertarget{gloss:rttime}{{\bf Round trip time:}}
Denoted by $R_{TT} = \sum_k R_k + Z$, it is the average time spent in the SUT together with the average time a virtual user spends thinking before issuing the next request. It can also be regarded as a special case of 
\hyperlink{gloss:rtime}{response time}.

%\noindent
\hypertarget{gloss:stime}{{\bf Service time:}}
Denoted by $S$, it is the average time spent in service, e.g., executing on a CPU or core. 
When there are repeated visits to the same resource for service, the cumulative time is called the service demand.
In the teletraffic context, $S$ is called the {\em holding time}. 
The first queue solved by Erlang was M/D/1 because he assumed a constant holding time~\cite{ERLA09}.

%\noindent
\hypertarget{gloss:ztime}{{\bf Think time:}}
Denoted by $Z$ it is the average delay (often associated with a human) between the completion of the previous request and the next.

%\noindent
\hypertarget{gloss:threads}{{\bf Threads:}}
Denoted by $N$, it represents the number of active load generation processes initiated by the driver-side  operating system. 

%\noindent
\hypertarget{gloss:wtime}{{\bf Waiting time:}}
Denoted by $W$, it is the mean time spent in waiting, e.g., in the scheduler run-queue, prior to receiving service. 
The terminology for waiting and queueing if often ambiguous across different but related disciplines, e.g., queueing theory vs. operating system architectures. 
In conventional queue-theoretic parlance~\cite{ERLA09,ERLA17,KLEI75,COOP84,GUNT11}, a queue is comprised of two parts: a service facility, which may contain multiple service resources, e.g., multiple cores, and a separate waiting line or buffer. 
In this sense, the ``run-queue'' in a scheduler is actually the waiting-line or run-buffer for \hyperlink{gloss:threads}{threads} that are ready to be executed, but not yet executing.

%\noindent
\hypertarget{gloss:uni}{{\bf Uniform distribution:}}
The uniform distribution (sometimes referred to as a Rectangular Distribution)
can be defined for either a discrete or a continuous random variable.
The continuous uniform distribution is the most applicable for load generation.
\begin{enumerate}
\item Continuous random variable defined on the real-valued interval $x \in [a,b]$
\item PMF $f(x) = \dfrac{1}{b-a}$ for $x \in [a,b]$ and zero elsewhere 
\item CDF $F(x) = \dfrac{x-a}{b-a}$ for $x \in [a,b)$, but zero for $x < a$ and 1 for $x \geq b$ 
\item Mean $E(x) = \frac{1}{2} (a+b)$
\item Variance $Var(x) = \frac{1}{12} (b-a)^2$
\item Standard deviation $SD = \frac{1}{\sqrt{12}} (b-a)$
\item Coefficient of variation $CoV = \dfrac{1}{\sqrt{3}} \bigg( \dfrac{b-a}{a+b} \bigg)$
\end{enumerate}

\end{document}